\documentclass[prc,floatfix,showpacs,amsmath,amsfonts]{revtex4}
\usepackage{amsmath,lscape,epsfig}
\usepackage{graphicx,color}

\def\beq{\begin{equation}}
\def\eeq{\end{equation}}
\def\beqa{\begin{eqnarray}}
\def\eeqa{\end{eqnarray}}
\def\ban{\begin{eqnarray*}}
\def\ean{\end{eqnarray*}}
\def\bi{\begin{itemize}}
\def\ei{\end{itemize}}

\newcommand{\fet}[1]{\mbox{\boldmath $#1$}}

\begin{document}

\title{Nucleon-nucleon scattering within a multiple subtractive renormalization approach }

\author{V. S. Tim\'oteo$^{1}$, T. Frederico$^{2}$, A. Delfino$^{3}$,}
\author{Lauro Tomio$^{3,4}$}
\affiliation{
$^{1}$ Faculdade de Tecnologia, Universidade Estadual de Campinas, 13484-332, Limeira, SP, Brazil.\\
$^{2}$ Instituto Tecnol\'ogico de Aeron\'autica, DCTA, 12228-900, S\~ao Jos\'e dos Campos, SP, Brazil.\\
$^{3}$ Instituto de F\'{\i}sica, Universidade Federal Fluminense, 24210-346, Niter\'oi, RJ, Brazil.\\
$^{4}$ Instituto de F\'{\i}sica Te\'orica, UNESP - Universidade Estadual Paulista, 
01140-070, S\~ao Paulo, SP, Brazil.}
\begin{abstract}
We present a methodology to renormalize the nucleon-nucleon
interaction in momentum space, using a recursive multiple
subtraction approach that prescinds from a cutoff regularization,
to construct the kernel of the scattering equation.
The subtracted scattering equation is solved with the next-leading-order 
(NLO) and next-to-next-leading-order (NNLO) interactions.  
The results are presented for all partial waves up to $j=2$, fitted 
to low-energy experimental data. 
In this renormalization group invariant approach, the subtraction 
energy emerges as a renormalization scale and the momentum associated 
with it comes to be about the QCD scale 
($\Lambda_{QCD}$), irrespectively to the partial wave.
\end{abstract}
\maketitle

\section{Introduction}

The nucleon-nucleon (NN) interaction in leading order  corresponds
to the one-pion-exchange potential (OPEP) plus a Dirac-delta, when
considering an effective field theory (EFT) of nuclear forces based
on a chiral expansion of the effective Lagrangian. This procedure
was suggested by Weinberg~\cite{wei}, with a recipe to infer the
values of the strength of the Dirac-delta interaction in the $^1S_0$
and $^3S_1$ channels from the singlet and triplet scattering lengths
respectively. Therefore, the singlet and triplet scattering lengths 
allows to fit the renormalized strengths of the contact interactions.
This effective potential should be valid for momenta well below some
typical momentum scale considered in quantum chromodynamics (QCD),
such as the rho meson mass ($m_\rho \sim$ 4 fm$^{-1}$)~\cite{wei},
which implies in a cutoff at the momentum scale of this order or
below, for the intermediate virtual propagation of the NN system.

According to  Weinberg's recipe, one should use naive dimensional 
analysis (NDA) to order terms in the potential, truncate it at a 
certain order, and then solve the Lippmann-Schwinger (LS)
equation exactly. However, it has been shown that this recipe is not
consistent with renormalization-group (RG) invariance, already at
leading order~\cite{wein-recipe,nogga05,va06}.
The validity of Weinberg's power counting is 
questioned~\cite{wein-recipe,nogga05,va06} 
in particular for the waves like $^3P_0$, where the singular and 
attractive pion tensor force is requiring more than two parameters 
even at LO. 

About a decade ago, an alternative way to renormalize the nucleon-nucleon 
interaction, for a singular potential was proposed in Ref.~\cite{npa99}.
In an extension of \cite{npa99}, the approach was proved in Ref.~\cite{plb00}
to be RG invariant, as it satisfies the corresponding non-relativistic 
Callan-Symanzik equations~\cite{CS}.
In this approach, no cutoff is considered in the equations and/or 
interactions. Instead, it is introduced a subtraction point in the 
kernel of the Lippmann-Schwinger (LS) equation~\cite{LS}, in order to 
reach a finite $T-$matrix.  
Our research group has considered singular contact interactions in the 
context of nuclear~\cite{fbbr-np,npa99,plb00,plb05}, atomic and general
physics~\cite{fbbr-ap,fbbr-gp}. In such works, within specific
renormalization procedures applied to few-body systems, scaling
limits and correlations between low-energy observables emerge as a
consequence of using singular contact (zero-range) interactions.

As the approach was proved to be invariant under renormalization group 
transformations, after fitting the data, one can easily move the reference 
scale without affecting the physics. This flexibility in moving
the reference scale is a big technical advantage of the present
approach in relation to other ones. Indeed, the renormalization
results for the observables should be independent of the specific
scheme used. Our method was applied to the NN interaction with
OPEP supplemented by contact interactions~\cite{npa99,plb00,plb05}.

The significant results obtained by several authors in the
implementation of the EFT program for the two-nucleon
system~\cite{ordonez,lepage,kaiser,friar,bira,epelrev06,epelbaum09,kaplan,birse,birse06,park,mehen,perry,machleidt,cohen,gegelia}
include a vast literature on the predictive power of the leading
order term (OPEP plus delta) with a single renormalization momentum
scale. Such calculation gave a sound basis for the renormalization
program of EFT in the NN system. In particular, the OPEP background
to the NN observables were carefully analyzed in Ref.~\cite{BR}.

Renormalization of the NN interaction in chiral effective theory in
leading order (LO), up to next-to-leading order (NLO), and to
next-to-next-to-leading order (NNLO), in S, P, D waves, has been
done with success in coupled and uncoupled waves
\cite{variolarev,variola09,entem,npa99,plb00,YEP08,YEP09,plb05}.
The references \cite{variolarev,variola09} apply the
renormalization approach in coordinate space, while the works
\cite{npa99,plb00,YEP08,YEP09,plb05} use subtracted-resolvent
two-body techniques in momentum space. The subtracted
renormalization procedure considered in Refs.~\cite{YEP08,YEP09} is
essentially different from the one applied in
Refs.~\cite{npa99,plb00,plb05} in the way to deal with the physical
inputs introduced to heal the ultraviolet divergences. In short: in
the renormalization strategy of \cite{YEP08}, one-folded subtracted
equation and also a cutoff are used, implying in a window where the
observables are quite independent on the cutoff. The strategy of
Refs.~\cite{npa99,plb00,plb05} is based on the elimination of
ultraviolet divergences and relies on multiple subtractions, without
any additional cutoff parameter.

The multiple subtraction technique that we are going to use in the
present approach demands a strategy to construct the driving term of
the corresponding scattering equation, which is generated by consecutive 
subtractions in the kernel at some defined energy scale. 
We denote a generic order of subtractions in the kernel 
by $n$, such that $n=1$ stands for one subtraction.
With the assumption that we have only $S-$wave LO contact interaction, 
one subtraction is enough. 
In this case the subtraction scale is
let to infinity and then driven towards a given finite value of the
subtraction energy by solving a non-relativistic Callan-Symanzik
(NRCS) equation~\cite{npa99,plb00}. The corresponding solution gives
the LO driven term calculated at the reference subtraction energy
where more contacts and TPE potentials are included.  
We introduce each order of the potential (LO, NLO, NNLO) in the driven
term of the subtracted equation when the number of subtractions are enough 
to make it finite.  For that, we use $n=3$ in case of NLO, and $n=4$ in NNLO.
Unitarity is strictly kept.
In this regard, we note that the lowest contact of the $P-$wave 
is introduced together with the TPEP NNLO.
In the complexity of this subject, to have different approaches to tame highly divergent potentials inherent to Weinberg proposal, it is important to observe 
the coherence between the physical results obtained with 
higher order potentials and the corresponding contacts to learn subtle 
aspects of QCD that permeates the chiral effective expansion of the nuclear
interaction. In other words, a trace of the QCD scale is expected to emerge
in the fitting procedure for the nucleon-nucleon scattering data. And,
as we are going to verify in our approach, this relic can be found in
the scale associated with the subtraction point where the two-pion
exchange potential at NLO and NNLO are introduced in the renormalization
process.

In the present work, going beyond the LO interaction and also 
considering $P-$ and $D-$partial waves, we show how to implement a 
multiple subtractive renormalization approach to obtain consistent 
results for the NN observables. We start by summarizing the methodology used to
renormalize the nucleon-nucleon interaction using a recursive
subtracted approach that was previously considered in
Ref.~\cite{npa99,plb00,plb05}. Next, we present the equations to be
solved in order to obtain a finite $T-$matrix with NLO and NNLO
interactions, taking into account partial waves up to $j=2$. The
results are fitted to low-energy experimental data. As it will be
shown, the adjusted derivative contact terms dominates over the NLO
and NNLO interactions in the $S-$wave channels for momentum higher
than about 1 fm$^{-1}$. It is about the same, in the case of nonzero
angular momentum. By including higher contact derivative terms, the
potential is fitted to low energy phase-shifts up to about 200 MeV
laboratory energies.

As we have verified, the subtractive-renormalization technique
considered in Refs.~\cite{npa99,plb00} is shown to be reliable also
for highly-singular potentials up to NNLO and $j=2$. These
potentials have also been discussed in Refs.~\cite{YEP09,YEP08}. In
order to show that, we present results for the nucleon-nucleon
phase-shifts up to $j=2$, going beyond previous calculations within
the present technique. We stress that our method cannot be naively
confused with a Born approximation which, in principle, is not
reliable for singular interactions. Indeed, we show that the
unitarity is strictly kept along all the subtractive procedure.
Moreover, it was already demonstrated in Ref.~\cite{plb00} that our
multiple subtractive approach is fully renormalization-group (RG)
invariant.

We also show how to implement the multiple subtractive
renormalization method for $P-$waves with contact interactions,
respecting the phase-shift threshold behavior of these waves. By
considering that, we supply the details not explicitly discussed in
Refs.~\cite{npa99,plb00}. 
Actually, $P-$wave channels were already
studied in Ref.~\cite{YEP09} with one subtraction and a sharp
momentum cutoff; and, in Ref.~\cite{variola09} (and references
therein), in coordinate space, even considering higher partial
waves.

The number of recursive steps required to renormalize the
interaction depends on how the potential diverges. For instance, the
leading order requires only one step, the next-to-leading order
requires three steps and the next-to-next-to-leading order requires
four steps.
The method has been extended for a generic number of
recursive steps \cite{plb00} and applied to one-pion exchange plus
contact interactions with three steps~\cite{plb05}. 
Our present renormalization technique can be used to organize and
implement calculations where one-pion exchange is treated
non-perturbatively. Higher order terms from the chiral expansion of the
nucleon-nucleon force can also be treated within this method.
A further advantage of our method can be particularly appreciated in 
the $^1S_0$ and $^3P_0$ channels, in which we show the phase-shift
behaviors for increasing cutoffs, exemplifying the cutoff independence   
as it goes to infinity.
In conception, our method differs from the works presented in 
Refs.~\cite{YEP08,YEP09} and \cite{variolarev,variola09} which also dealt 
with simpler effective chiral expansion in the way suggested by Weinberg.
In other approaches, such as the one given by Barford and
Birse~\cite{birse}, the higher chiral order are evaluated in
distorted-wave Born approximation. However, in this case the
analysis of the renormalization-group is quite involved. In
Ref.~\cite{epelrev06}, it was considered a systematic chiral
procedure with a smooth cutoff applied to the potential in order to
obtain the observables from the originally divergent scattering
equations. It has also been suggested that subleading orders of the
potential should not be fully iterated, but treated in
(distorted-wave) perturbation theory (see e.g. \cite{bira2008} and
\cite{valdepp}). 
 In our approach, we can iterate to all orders the potential and
still keep the renormalization group invariance.

This work is organized as follows. In section II, we briefly
describe the subtractive formalism applied to the Lippmann-Schwinger
equation, with the renormalization group equations. The formalism is
followed by an analytical example where the RG equation is solved
for the $P-$wave case with contact interactions, where two
subtractions are required. In section III, the renormalization
formalism is applied to the NNLO potential, by considering recursive
renormalization processes. The results for the NN phase-shifts are
presented and discussed in section IV. Our conclusions are given in
section V.

\section{Subtracted Lippmann-Schwinger Equation}

The scattering equation can be written with an arbitrary number of
subtractions $(n)$ in the kernel \cite{plb00,plb05}, that is useful
when the potential has ultraviolet divergences with attractive
nature e.g. $-1/r^m$ with $m\ge 2$ (for $m=2$ the critical strength
for collapsing the bound state is 1/4), and/or the potential
includes contact terms  (Dirac delta and it's derivatives).
By considering our units such that $\hbar$ and the nucleon mass
$m_N$ are equal 1, we have the energy $E$ given by $E\equiv k^2$.
Within such units, the regularized subtracted form of the LS equation,
in operatorial form, can be written as

\begin{widetext}
\begin{eqnarray}
T(E;-\mu^2)=
V^{(n)}(E;-\mu^2)\left[1 + G^{(n)}(E;-\mu^2) T(E;-\mu^2)\right]
&&\nonumber \\
=
\left[1 + T(E;-\mu^2) G^{(n)}(E;-\mu^2)\right]V^{(n)}(E;-\mu^2)
,&&
\label{LSn}
\end{eqnarray}
where, by having $H_0$ as the free operator, we have
\begin{eqnarray}
G^{(n)}(E;-\mu^2)&\equiv \left(\displaystyle
\frac{\mu^2+E}{\mu^2+H_0}\right)^nG^{(0)}(E), \label{Gn}
\end{eqnarray}
{\small
\begin{eqnarray} V^{(n)}(E;-\mu^2)&= V^{(n-1)}(E;-\mu^2)+V^{(n-1)}(E;-\mu^2)
(-\mu^2-E)^{n-1}\left[G^{(0)}(-\mu^2)\right]^n V^{(n)}(E;-\mu^2) \; .
\label{Vn1}
\end{eqnarray}
}
\end{widetext}
We should note that, $n$ is supposed to be the necessary maximum
number of subtractions, with $-\mu^2$ the energy scaling parameter,
to render finite results for the $T-$matrix. As we can verify in the
above equations, the dependence on the energy starts to appear in
$V^{(n)}$ for $n\ge 2$.
The formal solution of the regularized Eq.~(\ref{LSn}),
for $V^{(n)}\equiv V^{(n)}(E;-\mu^2)$, is given by
\begin{small}
\begin{eqnarray}
T(E;-\mu^2)
&=& \left[1- V^{(n)} G^{(n)}(E;-\mu^2)\right]^{-1} V^{(n)}
\nonumber\\
&=& V^{(n)}\left[1- G^{(n)}(E;-\mu^2)V^{(n)} \right]^{-1}
\; .  \label{VnI}
\end{eqnarray}
\end{small}
By considering the operatorial expression given by Eq.~(\ref{Vn1})
with $n$ subtraction, in explicit momentum-space notation, for an
arbitrary single partial-wave (after integrated the angular parts
and with the assumption that the
interaction is symmetric and not angle dependent), and by
taking $V^{(n)}(q,p;k^2;-\mu^2)$ as the matrix element of the
angular momentum projected operator, with
\begin{eqnarray}
\frac{2}{\pi}\int_0^{\infty}dp~p^2\;\; |p\rangle \langle p| &\equiv&
{\bf 1},
\end{eqnarray}
we obtain the partial-wave projected equation, which is given by
\begin{small} \begin{widetext}
\begin{eqnarray}
V^{(n)}(q,p;k^2;-\mu^2)&=& V^{(n-1)}(q,p;k^2;-\mu^2)
\nonumber\\ &-&
\frac{2}{\pi}\int_0^{\infty}dq'~q'^2~V^{(n-1)}(q,q';k^2;-\mu^2)
\frac{(\mu^2+k^2)^{n-1}}{(\mu^2+q'^2)^n} ~V^{(n)}(q',p;k^2;-\mu^2)
\; . \label{Vn}
\end{eqnarray}
\end{widetext} \end{small}
Therefore, as an example, we can evolve
in subtractions the OPEP plus delta, which is recognized to be
renormalizable by fixing only one $S-$wave observable in the coupled
triplet and singlet states (usually the scattering length is chosen
as the input for fixing the renormalized strength of the contact).
In this particular case, when only one subtraction is enough to
produce a finite $T$-matrix, $V^{(1)}(-\mu^2)$ is replaced by the
corresponding $T-$matrix at the point $E=-\mu^2$, such that,
by defining $T(-\mu^2;-\mu^2)\equiv T(-\mu^2)$, we can obtain the
following subtracted equation:
\begin{eqnarray}\label{SKLS}
&&T(E;-\mu^2)=
T(-\mu^2)\left\{1+
\left[ G^{(0)}(E) - G^{(0)}(-\mu^2) \right]~T(E;-\mu^2)\right\} .
\nonumber
\end{eqnarray}
For the Dirac delta-potential, the matrix element in momentum space
of $T(-\mu^2,-\mu^2)$ is the renormalized strength.

The subtraction energy $-\mu^2$ in Eq.~(\ref{LSn}) can be moved
without changing the resulting $T-$matrix if the driving term is
evolved by solving the Callan-Symanzik equation~\cite{CS,plb00,plb05}.
The renormalization group equation
for the driving term is briefly sketched below. One example of
renormalization of a $P-$wave case, with our method, is also
discussed for illustration. On the calculations of shallow $P-$wave
states, with EFT theory applied for halo nuclei, see
Ref.~\cite{bertu-bira}.

In addition, we remark a general feature of the subtraction
procedure applied to attractive and repulsive power-law potentials
in configuration space ($r^{-m}$ with  $m\ge 2$). In the repulsive
case the subtraction is not trivially required to render finite the
$T-$matrix, as the scattering wave function is damped in the
classically forbidden region. In contrast, in the attractive case
the ultraviolet divergence is actually met dynamically and may
collapse the state. That is the case that a regularization has to be
performed at the short range. In our method we deal with this
problem, by using enough subtractions to make the $T-$matrix finite.
Attractive and repulsive terms that we are discussing now appears
as the chiral expansion is performed, as also discussed thoroughly by 
Valderrama and Arriola in their renormalization scheme in configuration space
for calculation of the NN phase-shifts \cite{variolarev}. Instead,
in our approach, subtractions are introduced to include the contacts,
at the expense of finding a convenient subtraction point with the meaning of
a reference low-energy scale. As we will show, indeed the fitting of
partial waves phase-shifts up to $j=2$ gives a reference subtraction
energy between -50 and -100 MeV. At this scale the NLO and NNLO TPEP
potentials are identified as contributions to  the driving terms as
we are going to explain in detail in Sect. III.

Note that the reference subtraction point is, in principle,
arbitrary, and can be moved by evolving the potential through the RG
flow equation.
Once the fitting to the data is done,  the reference scale can be
changed arbitrarily as long as the driving term (starting potential)
is evolved through the nonrelativistic Callan-Symanzik flow
equation~\cite{CS}.
In fact, this implies that the RG invariance cannot be separated from
the form of the evolved interaction.
Next, we briefly discuss the RG equations, followed by an analytical example.

\subsection{Renormalization Group Equations}

The nucleon-nucleon observables are invariant under the change of
the arbitrary subtraction point, therefore one can start at any
convenient energy scale $-\mu^2$.
However, the form of the driving term and its coefficients,
which define the scattering amplitude are tied to the prescription used to
define the renormalized theory. The key point of the
renormalization group method is to change this prescription
without altering the predictions of the theory~\cite{weinbook}.

The invariance of the $T-$matrix under changes of renormalization
prescriptions, i.e.,
\begin{eqnarray}
\frac{\partial}{\partial\mu^2} T(E;-\mu^2)=0 \ ,
\end{eqnarray}
 imposes a definite rule to modify $V^{(n)}$
that appears in a form of a non-relativistic Callan-Symanzik (NRCS) equation~\cite{plb00,CS}:
\begin{small}
\begin{eqnarray}
&&\frac{\partial V^{(n)}(E;-{ \mu}^2)}{\partial \mu^2} =\nonumber\\
&=&-V^{(n)}(E;-{ \mu}^2) \frac{\partial G^{(n)}(E;-{\mu}^2)
}{\partial \mu^2} V^{(n)}(E;-{ \mu}^2) \  \label{ss3}\\
&=& n\;
V^{(n)}(E;-{ \mu}^2)
\frac{\left(E+{ \mu}^2\right)^{n-1}}{\left(H_0+{ \mu}^2\right)^{n+1}
}
V^{(n)}(E;-{ \mu}^2) \ ,
\label{dgn}
\end{eqnarray}
\end{small}
which is derived from the invariance of $T(E;-{ \mu}^2)$, given
by Eq.~(\ref{VnI}), with respect to the renormalization parameter $\mu^2$.
The demonstration of the above can easily be done by considering
both expressions that appear in Eq.(\ref{VnI}).
Equation~(\ref{ss3}) substantiate the invariance of the renormalized
$T-$matrix under dislocation of the subtraction point. Then, we
observe that there is a non-trivial dependence on the subtraction
point appearing in the driving term of the subtracted scattering
equation, although the physical results of the model are kept
unchanged.  Thus, from now on, we drop the explicit dependence on
$\mu^2$ in the T-matrix by writing $T(E;-\mu^2)\equiv T(E)$.

The solution of (\ref{ss3}) implies in a complicated evolution of
$V^{(n)}$ as $\mu$ changes. Not only the strengths of the
interactions would change, but also the form of the driving term.
The ultraviolet behavior of the driving term is not changed by the
evolution in $\mu$. The evolution should not be truncated as $\mu$
is varied in order to keep the $T-$matrix invariant. At different
$\mu$ the potential $V^{(n)}(-\mu^2;E)$ has a complicate form from
the solution of NRCS equation. Similarly, the evolution of
renormalization group equations as introduced by Bogner, Kuo and
Schwenk~\cite{bogner} for the NN scattering does not truncate on
certain operators as the cutoff is varied  to keep the observables
unchanged.

\subsection{Example: Subtracted $P-$wave equations with contact interactions}

We illustrate the use of our multiple subtraction renormalization
method by discussing the case of a one-channel $P-$wave problem with
a derivative contact interaction
$V(p,p^\prime)=\lambda_1~p~p^\prime$. The $P-$wave
Lippmann-Schwinger equation with a momentum cutoff $\Lambda$ is easily
solvable, with the corresponding $T-$matrix given by
\begin{small}\begin{eqnarray}
T(p,p^\prime;k^2)=p\cdot
p^\prime\left(\frac{1}{\lambda_1}-\frac{2}{\pi}\int_0^\Lambda dq \;
q^4\frac{1}{k^2-q^2+{\rm i} \epsilon}\right)^{-1} \ . \label{p1}
\end{eqnarray}\end{small}
The coefficient $\lambda_1$ can be fixed by the scattering volume $\alpha$,
which gives
\begin{eqnarray}
\frac{1}{ \lambda_1} =-\frac{2}{3\pi}\Lambda^3+\frac{1}{\alpha}.
\end{eqnarray}
By replacing $\lambda_1$ in Eq.~(\ref{p1}), we have
\begin{small}\begin{eqnarray}
T(p,p^\prime;k^2)=p\cdot
p^\prime\left(\frac{1}{\alpha}-\frac{2}{\pi}k^2\int_0^\Lambda dq \;
q^2\frac{1}{k^2-q^2+{\rm i} \epsilon}\right)^{-1} \ , \label{p2}
\end{eqnarray}\end{small}
which shows that this procedure is not enough to eliminate the cutoff dependence of Eq.~(\ref{p1}).
With this simple example, we can see that a renormalization procedure 
using only one subtraction, for a $P-$wave problem, is expected to be 
cutoff dependent, requiring a second subtraction.

The fit of only the scattering volume is equivalent in our method to
use one subtraction in the Lippmann-Schwinger equation. One should
note that for one subtraction $n=1$, the dependence on the cutoff is
linear. 
Even in this simple example, the renormalization group evolution of the 
recursive driving term is not straightforward.  But it is important to 
stress that one more parameter is required to renormalize Eq.~(\ref{p2}).
Therefore, in agreement with Refs.~\cite{bertu-bira,harada}, to 
renormalize the $P-$wave contact interaction we need two
inputs: the scattering volume and the next term in the generalized
effective range expansion for the $k^3 \cot \delta$. 

Within our method, with a scaling momentum parameter $\mu$, instead of
the cutoff $\Lambda$, the ultraviolet divergence in the scattering
equation from the potential  $V(p,p^\prime)=\lambda_1\;p\;p^\prime$
requires two subtractions  $(n=2)$ to render finite the integral appearing
in Eq.~(\ref{p1}). The corresponding result, where we can take the limit
$\Lambda \to \infty$, is given by
\begin{widetext}
\begin{eqnarray}
&&\frac{T(p,p^\prime;k^2)}{p~p^\prime}=
\left(\frac{1}{\lambda_{1}(k^2,\mu^2)}
-\frac{2}{\pi} \int_0^\infty dq
\frac{q^4(k^2+\mu^2)^2}{\left(\mu^2+q^2\right)^2\left(k^2-q^2+{\rm i}
\epsilon\right)}\right)^{-1} , \label{p3}
\end{eqnarray}
\end{widetext}
where the parameter $\lambda_{1}$ is now finite and can be fitted by the
scattering volume. However, a dependence on $\mu$ is intrinsic to this method,
unless the driving term is evolved through a nonrelativistic Callan-Symanzik
equation~\cite{plb00,plb05}.

Just to give an impression on how the RG equation works for the $P-$wave
example, we introduce the renormalized driving term
$V(p,p^\prime)\equiv V(p,p^\prime;k^2)$ $=\lambda_{1}(k^2,\mu^2)~p~p^\prime$ 
in Eq.~(\ref{dgn}). After projection in the $P-$wave, we have
\begin{widetext}
\begin{eqnarray}
\frac{\partial \lambda_{1}(k^2,\mu^2)}{\partial \mu^2}
&=&
2\left[\lambda_{1}(k^2,\mu^2)\right]^2 \frac{2}{\pi} \int^\infty_0
dq\; q^4\frac{(k^2+\mu^2)}{(q^2+\mu^2)^3} 
= \frac{3}{4} \frac{(k^2+\mu^2)}{\mu}
\left[\lambda_{1}(k^2,\mu^2)\right]^2 \ . \label{ss4}
\end{eqnarray}
\end{widetext}
Therefore, the solution of the RG equation gives
{\small
\begin{eqnarray}
\frac{1}{\lambda_{1}(k^2,\mu^2)}=\frac{1}{\lambda_{1}(k^2,\mu_0^2)} - \frac32 k^2(\mu-\mu_0)
- \frac12(\mu^3-\mu^3_0).
\label{ss5}
\end{eqnarray}
}
Note that, even if at the reference subtraction energy $-\mu^2_0$
the renormalized strength is chosen to be independent of $k^2$, the
evolution introduces a $k^2$ dependence, and implicitly the need of
two-subtractions to render finite the T-matrix for the $P-$wave. If
one fixes the reference scale at zero energy as suggested in 
\cite{YEP09}, the renormalized strength is the scattering volume
assuming independence with $k^2$ at this reference energy.
In our case, after solving the RG equation, the form of the
evolved potential is simply the renormalized $\lambda_1$ from
Eq. (\ref{ss5}) times $p~p'$. Solving Eq. (\ref{p3}) with the renormalized
coupling from Eq. (\ref{ss5}), we get
\begin{eqnarray}
k^3 \cot \delta = -\frac{1}{\lambda_1(k^2,\mu^2)} - \frac{\mu}{2}
(3k^2+\mu^2).
\label{k3cot}
\end{eqnarray}
The evolution of $\lambda_1$ according to Eq. (\ref{ss5}) shows
that it should be linear function of $k^2$, which demands two constants.
Indeed, if we re-write Eq. (\ref{k3cot}) taking into account the constant plus
the $k^2$ term from $\lambda_1$ only two independent quantities appear, i.e.,
a scattering volume and an effective momentum.
In fact, the form we obtain for $k^3\cot\delta$ is similar to the one obtained
in Ref.~\cite{bertu-bira}.

\section{Renormalization of the NNLO Potential with $n=4$}

Once we have established the recursive procedure to renormalize the
$NN$ interaction, we need a potential in momentum space.
For the NNLO chiral potential, we adopt a momentum-space form as explicitly 
given by Epelbaum~\cite{epelrev06}.

Even though our method is powerful enough to renormalize the full two-pion
exchange (TPE) potential, we consider the version with the spectral representation
regularization such that the comparison with results obtained by other calculations
can be more straight.

For the sake of completeness and the reader's convenience, we repeat
here the analytical expressions for the chiral NNLO momentum space potential.
The LO interaction is given by the one-pion exchange (OPE) plus an
contact interaction. The strength of the isospin conserving contact terms and their
derivatives depends on the LSJ channel, considering also angular momentum mixing.
We simplify our notation for the contact terms so that the total angular momentum and
isospin dependence are not shown explicitly in the expressions for the potential.
The corresponding $S-$wave projected matrix elements of the interaction is given by
\begin{equation}
V_{LO}(p,p') = V_{\rm OPE}({p},{p'}) + \lambda_0 \; , \label{11}
\end{equation}
where the unprojected OPE potential, for $\vec{q}\equiv \vec{p}-\vec{p'}$,
is given by
{\small \begin{eqnarray}
V_{\rm OPE}(\vec{p},\vec{p'}) =
\frac{-1}{(2\pi)^3}\left(
\frac{g_A}{2f_\pi}\right)^2
\fet{\tau}_1 . \fet{\tau}_2 \, \frac{(\vec{\sigma}_1 \cdot\vec{q})\,(\vec{\sigma}_2\cdot\vec{q})}
{q^2 + M_\pi^2}~,
\end{eqnarray} }
At NLO, we have some TPE diagrams plus derivative contact interactions.
After partial wave projection, it is given by
\begin{eqnarray}
V_{NLO}(p,p') &=& V^{NLO}_{\rm TPE}(p,p')
+\lambda_1~(p~p')\delta_{L,1}\delta_{L',1}\nonumber\\
&+&\left(\lambda_2~(p^2+{p'}^2)
+\lambda_3~(p^2~{p'}^2)\right)\delta_{L,0}\delta_{L',0}\nonumber\\
&+&
\lambda_4~\left(p^2\delta_{L,2}\delta_{L',0}+
{p'}^2\delta_{L',2}\delta_{L,0}\right) \; , \label{13}
\end{eqnarray}
where the unprojected NLO TPE potential is
{\small
\begin{widetext}
\begin{eqnarray}
V^{NLO}_{\rm TPE}(\vec{p},\vec{p'})
&=& - \left( \frac{ \fet{\tau}_1 \cdot \fet{\tau}_2 }{384 \pi^2 f_\pi^4} \right) \,
\frac{L(q)}{(2\pi)^3} \, \biggl\{4M_\pi^2 (5g_A^4 - 4g_A^2 -1) + q^2(23g_A^4 - 10g_A^2 -1)
+ \frac{48 g_A^4 M_\pi^4}{4 M_\pi^2 + q^2} 
\biggr\} \nonumber\\ &&  
- \left( \frac{3 g_A^4}{64 \pi^2 f_\pi^4} \right)  \,L(q)  \, \biggl\{
(\vec{\sigma}_1 \cdot\vec{q})\,(\vec{\sigma}_2\cdot\vec{q}) - q^2 \,
\vec{\sigma}_1 \cdot\vec{\sigma}_2 \biggr\} ~.
\end{eqnarray}
\end{widetext}
}
The term proportional to $\lambda_1$ contributes only in the
$P-$waves, the terms proportional to $\lambda_2$ and $\lambda_3$
appear only in the $S-$waves and the the term proportional to
$\lambda_4$ enters only in the coupled channels with $j=1$.
The $\lambda_3$ term ($p^2~p'^2$) actually appears at N$^3$LO
in the Weinberg's power counting, but we promoted it to NLO in order
to improve the $S-$waves fit. It could have been promoted to NNLO, but
since the iteration of the NLO $\lambda_2$ term $(p^2 + p'^2)$ leads
to $p^2~p'^2$ terms, we included it at NLO so that all the
$p^2~p'^2$ terms combine at once. This has also been done in Ref.
\cite{plb05}.

Finally, at NNLO, we have other TPE diagrams,
with the corresponding unprojected potential given by
{\small
\begin{widetext}
\begin{eqnarray}
&&V_{NNLO}(\vec{p},\vec{p'})\equiv V^{NNLO}_{\rm TPE}(\vec{p},\vec{p'}) \nonumber\\
&=&\frac{1}{(2\pi)^3} \left( \frac{3g_A^2}{16\pi f_\pi^4} \right) \biggl\{
\frac{g_A^2 M_\pi^5}{16 m (4M_\pi^2+q^2)} - \biggl[2M_\pi^2(2c_1 -c_3) -q^2 \,
\bigl( c_3 + \frac{3g_A^2}{16m} \bigr)\biggr] (2M_\pi^2+q^2) A(q) \biggr\}
\nonumber\\
&-& \frac{g_A^2}{128\pi m f_\pi^4} (\fet{ \tau}_1 \cdot \fet{ \tau}_2 ) \,
\biggl\{ -\frac{3g_A^2  M_\pi^5}{4M_\pi^2+q^2} + \bigl[ 4M_\pi^2 +
2q^2 -g_A^2 (4M_\pi^2 + 3q^2) \bigr]  (2M_\pi^2+q^2) A(q) \biggr\}
\nonumber\\
&+&  \frac{9g_A^4}{512\pi m f_\pi^4} \biggl[ (\vec \sigma_1 \cdot \vec
q\,)(\vec \sigma_2 \cdot \vec q\,) -q^2 (\vec \sigma_1 \cdot\vec \sigma_2
)\biggr] \, (2M_\pi^2+q^2) A(q)  \nonumber\\
&-& \frac{g_A^2}{32\pi f_\pi^4} (\fet{ \tau}_1 \cdot \fet{ \tau}_2 ) \,
\biggl[ (\vec \sigma_1 \cdot \vec q\,)(\vec \sigma_2 \cdot \vec q\,)
-q^2 (\vec \sigma_1 \cdot\vec \sigma_2 )\biggr]
\biggl\{ \bigl( c_4 + \frac{1}{4m} \bigr) (4M_\pi^2 + q^2)
-\frac{g_A^2}{8m} (10M_\pi^2 + 3q^2) \biggr\} \, A(q) \nonumber\\
&-& \frac{3g_A^4}{64\pi m f_\pi^4} \, i \, (\vec \sigma_1 +  \vec
\sigma_2 ) \cdot (\vec{p}~' \times \vec{p} ) \, (2M_\pi^2+q^2) A(q) \nonumber\\
&-& \frac{g_A^2(1-g_A^2)}{64\pi m f_\pi^4} (\fet{ \tau}_1 \cdot \fet{ \tau}_2 )
\, i \, (\vec \sigma_1 +  \vec \sigma_2 ) \cdot (\vec{p}~' \times
\vec{p} ) \, (4M_\pi^2+q^2) A(q) \; ,
\end{eqnarray}
\end{widetext}
}
where the loop integrals $L(q)$ and $A(q)$ are given by
\begin{eqnarray}
L(q) &=& \frac{1}{q}\sqrt{4 M_\pi^2 + q^2}
\ln\frac{\sqrt{4 M_\pi^2 + q^2}+q}{2M_\pi}~, 
\nonumber \\ &\approx&  
\theta(\tilde\Lambda-2M_\pi) \frac{\sqrt{4 M_\pi^2 + q^2}}{2q}
\ln\frac{\left(\tilde\Lambda\sqrt{4 M_\pi^2 + q^2}+q\sqrt{{\tilde\Lambda}^2-4 M_\pi^2}\right)^2}
{4 M_\pi^2({\tilde\Lambda}^2-q^2)}
~, \nonumber \\
A(q) &=& \frac{1}{2q} \arctan \frac{q}{2M_\pi} ~,
\approx \theta(\tilde\Lambda-2M_\pi)
\frac{1}{2q} \arctan\frac{q(\tilde\Lambda-2M_\pi)}{q^2+2M_\pi\tilde\Lambda}~.
\end{eqnarray}
In the above expressions, ${\tilde\Lambda}$ 
is a spectral regularization scale defined in \cite{epelrev06} for the 
two-pion exchange, in order to improve the convergence of the chiral expansion.
As suggested ${\tilde\Lambda}\approx $ four pion mass, we set ${\tilde\Lambda}=$ 600 MeV.

\subsection{Evolving the potentials through the recursive renormalization process}

We start by calculating $V^{(1)}(-\mu^2)$ from the leading order
interaction $V_{LO}$, by  solving the Callan-Symanzik equation
(\ref{ss3}) for $n=1$ starting at a negative infinite $\bar\mu^2$ up
to a finite $\mu^2$. The integral form of RG equation (\ref{ss3})
for $n=1$, after partial wave decomposition, is
{\small 
\begin{eqnarray}
V^{(1)}(p,p';-\mu^2)
&=& V_{LO}(p,p';-\bar\mu^2)
+ \frac{2}{\pi} \int_0^{\infty} dq ~ q^2 ~\frac{V_{LO}(p,q;-\bar\mu^2) (\mu^2-\bar\mu^2)}{(\bar\mu^2+q^2)(\mu^2+q^2)}V^{(1)}(q,p';-\mu^2)  \; ,
\label{v1ppl}
\end{eqnarray}
}
which brings the leading order interaction to a scale $-\mu^2$
from its infinitely large fixed-point
$-\bar\mu^2$~\cite{birse,birse06,npa99,plb05}.
This generates an interaction for the one subtracted scattering
equation, which gives the same observables as the leading order
interaction when $\mu \gg \Lambda_{QCD}$.  
Now, we obtain $V^{(2)}(p,p';k^2;-\mu^2)$ from
$V^{(1)}(p,p';-\mu^2)$:
\begin{widetext}
{\small
\begin{eqnarray}
V^{(2)}(p,p';k^2;-\mu^2) = V^{(1)}(p,p';-\mu^2)
-  \frac{2}{\pi} \int_0^{\infty} dq ~ {q}^2 ~\frac{V^{(1)}(p,q;-\mu^2)(\mu^2+k^2)}{(\mu^2+q^2)^2}
V^{(2)}(q,p';k^2;-\mu^2) \; .\label{v2ppl}
\end{eqnarray}
}
At the third step we evolve from $V^{(2)}$ and introduce the next-to-leading order terms
{\small
\begin{eqnarray}
\bar{V}^{(3)}(p,p';k^2;-\mu^2) &=&  V^{(2)}(p,p';k^2;-\mu^2)\nonumber\\
&-& \frac{2}{\pi} \int_0^{\infty} dq ~ {q}^2 ~\frac{V^{(2)}(p,q;k^2;-\mu^2)(\mu^2+k^2)^2}{(\mu^2+q^2)^3}\bar{V}^{(3)}(q,p';k^2;-\mu^2) \; , \nonumber \\
V^{(3)}(p,p';k^2;-\mu^2)~ &=& ~V_{NLO}(p,p';-\mu^2)
+ \bar{V}^{(3)}(p,p';k^2;-\mu^2) \; . \label{v3+ppl}
\end{eqnarray}
}
At the fourth step, the higher order we consider here, we evolve from $V^{(3)}$ and
add the next-to-next-to-leading order two-pion exchange:
\begin{eqnarray}
\bar{V}^{(4)}(p,p';k^2;-\mu^2) &=& V^{(3)}(p,p';k^2;-\mu^2)\nonumber\\
&-&  \frac{2}{\pi} \int_0^{\infty} dq ~ {q}^2 ~V^{(3)}(p,q;k^2;-\mu^2)\frac{(\mu^2+k^2)^3}{(\mu^2+q^2)^4}\bar{V}^{(4)}(q,p';k^2;-\mu^2)   \nonumber\\
V^{(4)}(p,p';k^2;-\mu^2)~ &=&~V_{NNLO}(p,p';-\mu^2) + \bar{V}^{(4)}(p,p';k^2;-\mu^2) \; .\label{v4+ppl}
\end{eqnarray}
With the above, the half-on-shell $T-$matrix with four subtractions is
a solution of
\begin{eqnarray}
T(p,k;k^2) &=& V^{(4)}(p,k;k^2;-\mu^2) \nonumber\\
&+&
\frac{2}{\pi} \int_0^{\infty}
dp' ~{p'}^2 ~
V^{(4)}(p,p';k^2;-\mu^2)\left(\frac{\mu^2+k^2}{\mu^2+p'^2}\right)^4
\frac{1}{k^2-p'^2+{\rm i}\epsilon}~T(p',k;k^2) \; \label{thon}.
\end{eqnarray}
\end{widetext}
Note that in the above equation the term given by
$\left[(\mu^2+k^2)/(\mu^2+p'^2)\right]^4$ works effectively as a
regulator, canceling the singularity presented in the starting
interaction. The driving term is generated by consecutive
subtractions in the kernel at some defined energy scale $\mu^2$,
with the advantage that such scale can be moved freely as long as
$V^{(4)}$ satisfies the RG equation, given by Eq.~(\ref{ss3}), that guarantees
that the scattering amplitude is unaltered.

It is important to explain the strategy used  to insert the
potential in the recursive to obtain the driving term $V^{(4)}$ that
enters in Eq.~(\ref{thon}). We include each order of the interaction
(LO, NLO, NNLO) in the step where there is enough subtractions to
renormalize it. Consequently, we insert the LO potential in the
first subtraction, the NLO potential in the third subtraction and so
on. Tables \ref{tab1} to \ref{tab4} list the parameters for the potentials used in
this work. 

Before closing this subsection, we discuss briefly our
scheme and the one used in Ref. \cite{YEP09}, where $S-$waves have
been considered with more than one subtraction. In that work,  the
successive subtractions are performed in a rather different form
compared to our approach. They call it a mixed scheme. The main
difference is: in order to handle the momentum-dependent contact
interaction $p^2 + p'^2$, the cutoff also plays an important role in
\cite{YEP09}, since according to our method this interaction
actually requires three subtractions. In our case, we can approach
any kind of interaction as long as we perform enough subtractions.
The cutoff has no physical relevance since the multiple subtractions
allow us to sum up to arbitrarily large momentum values. 

\begin{table}[ht]
\begin{center}
\caption{Strengths of the LO contact interactions, which reproduce the
scattering lengths for the $S$ waves. The values of $\lambda_0^{^1S_0}$ and $\lambda_0^{^3S_1}$, 
in units of fm, are given at the energy scale $-\bar\mu^2$, with $\bar\mu=$ 30 fm$^{-1}$
($\bar\mu^2=41.47\times$ 900 MeV). } \label{tab1}
\begin{tabular}{c|cc}
\hline \hline
       Strengths      & $^1S_0$ &     $^3S_1$\\
 \hline\hline
$\lambda_0$(fm)       & -0.0203 &   -0.24142 \\
\hline \hline
\end{tabular}
\end{center}
\begin{center}
\caption{Strengths of the contact interactions 
for the fits with the LO potential plus the NLO contact interactions. 
The values of $\lambda_0^{^1S_0}$ and $\lambda_0^{^3S_1}$ are given 
at the same energy scale as in Table \ref{tab1} ($-\bar\mu^2=-41.47\times$ 900 MeV); 
with $\lambda_2^{^1S_0}$ and $\lambda_3^{^1S_0}$ at $-\mu^2=$ -50 MeV. The other 
strengths are given at $-\mu^2=$-100 MeV. 
} \label{tab2}
\begin{tabular}{c|ccccccc}
\hline \hline
Strengths            &$^1S_0$ & $^3P_0$ &  $^3S_1$&   $^1P_1$   &    $^3P_1$ &   $^3P_2$&    $\epsilon_1$  \\
\hline \hline
$\lambda_0$(fm)      & -0.0165&      -  & -0.2480 &     -       &       -    &    -     &          -     \\
$\lambda_1$(fm$^3$)  &   -    & 0.25    & -       &     0.04    &     0.007  &   -0.07  &         -     \\
$\lambda_2$(fm$^3$)  & 2.2660 &      -  &      0.1&      -      &       -    &     -    &          -     \\
$\lambda_3$(fm$^5$)  & 2.0047 &      -  &       - &      -      &       -    &     -    &          -     \\
$\lambda_4$(fm$^3$)  & -      &      -  &       - &      -      &       -    &     -    &         0.001 \\
\hline \hline
\end{tabular}
\end{center}
\caption{Strengths of the contact interactions for the 
fits with the full NLO potential. The values of the
$\lambda$'s are given for $\bar\mu^2$ and $\mu^2$ as in Table \ref{tab2}.}
\label{tab3}
\begin{center}
\begin{tabular}{c|ccccccc}
\hline \hline
Strengths          &$^1S_0$ &$^3P_0$ &$^3S_1$  & $^1P_1$    &  $^3P_1$  & $^3P_2$  & $\epsilon_1$  \\
\hline\hline
$\lambda_0$        &-0.0190 &      - & -0.1602 &     -      &      -    &    -     &      -        \\
$\lambda_1$(fm$^3$)&   -    & 0.37   & -       &     0.063  & -0.078    & -0.04    &      -        \\
$\lambda_2$(fm$^3$)&2.2660  &     -  & 0.1     &     -      &    -      &    -     &      -        \\
$\lambda_3$(fm$^5$)&2.0047  &      - & -       &     -      &    -      &    -     &      -        \\
$\lambda_4$(fm$^3$)&-       &      - & -       &     -      &    -      &    -     & 0.17          \\
\hline \hline
\end{tabular}
\end{center}
\caption{Strengths of the contact interactions for the fits with the NNLO potential.  
The values of the $\lambda$'s are given for $\bar\mu^2$ and $\mu^2$ as in Table 
\ref{tab2}.}
\label{tab4}
\begin{center}
\begin{tabular}{c|ccccccc}
\hline \hline
Strengths          & $^1S_0$ & $^3P_0$ & $^3S_1$ & $^1P_1$ & $^3P_1$   &   $^3P_2$  &    $\epsilon_1$  \\
\hline\hline
$\lambda_0$(fm)    & -0.0189 & -       &-0.1217  &   -     &   -       &     -      &         -         \\
$\lambda_1$(fm$^3$)& -       & 0.303   & -       & 0.066   & -0.19     &   -0.1     &         -         \\
$\lambda_2$(fm$^3$)& 2.2660  & -       & 0.1     &   -     &   -       &     -      &         -        \\
$\lambda_3$(fm$^5$)& 2.0047  & -       & -       &   -     &   -       &     -      &         -        \\
$\lambda_4$(fm$^3$)& -       & -       & -       &   -     &   -       &     -      &       0.17            \\
\hline \hline
\end{tabular}
\end{center}
\end{table}

\subsection{Half-on-shell amplitudes for the recursive renormalization process}

Now, let us consider the calculation of the half-on shell matrices
$V^{(n)}(q,k)$ for $n=$1 to 4, from the solution of (\ref{v1ppl})
to (\ref{v4+ppl}), and the half-on-shell matrix elements of
$T(q,k;k^2)$ solution of (\ref{thon}). Before starting our
discussion, a side remark is worthwhile. As we have seen before,
from the above equations, the recursive driving terms obey integral
equations with kernels defined by subtracted free Green's functions
multiplied by recursive driving terms one order below. From the
subtracted method itself the lowest order recursive driving term
satisfies an equation which is energy independent. In the following
orders the integral equations for the driving term depend on $k^2$,
and the subtracted Green's function is negative, which brings a
curious effect on the respective solutions of (\ref{Vn}). The
corresponding homogeneous equation, in the case of attractive
recursive interaction, suffers an enhancement by increasing $k^2$,
due to the factor $(\mu^2+k^2)$ inside the kernel, which may allow
it to have a solution. When this happens, an unphysical pole will
occur in the recursive interaction of the integral equation. This
can be realized even with a regular attractive potential if multiple
subtractions are used to compute the scattering amplitude. This
unphysical pole is completely washed out in the solution of the
$n-$subtracted LS equation given below.

\begin{figure}[t]
\centering
\includegraphics[height=10cm, width=12cm]{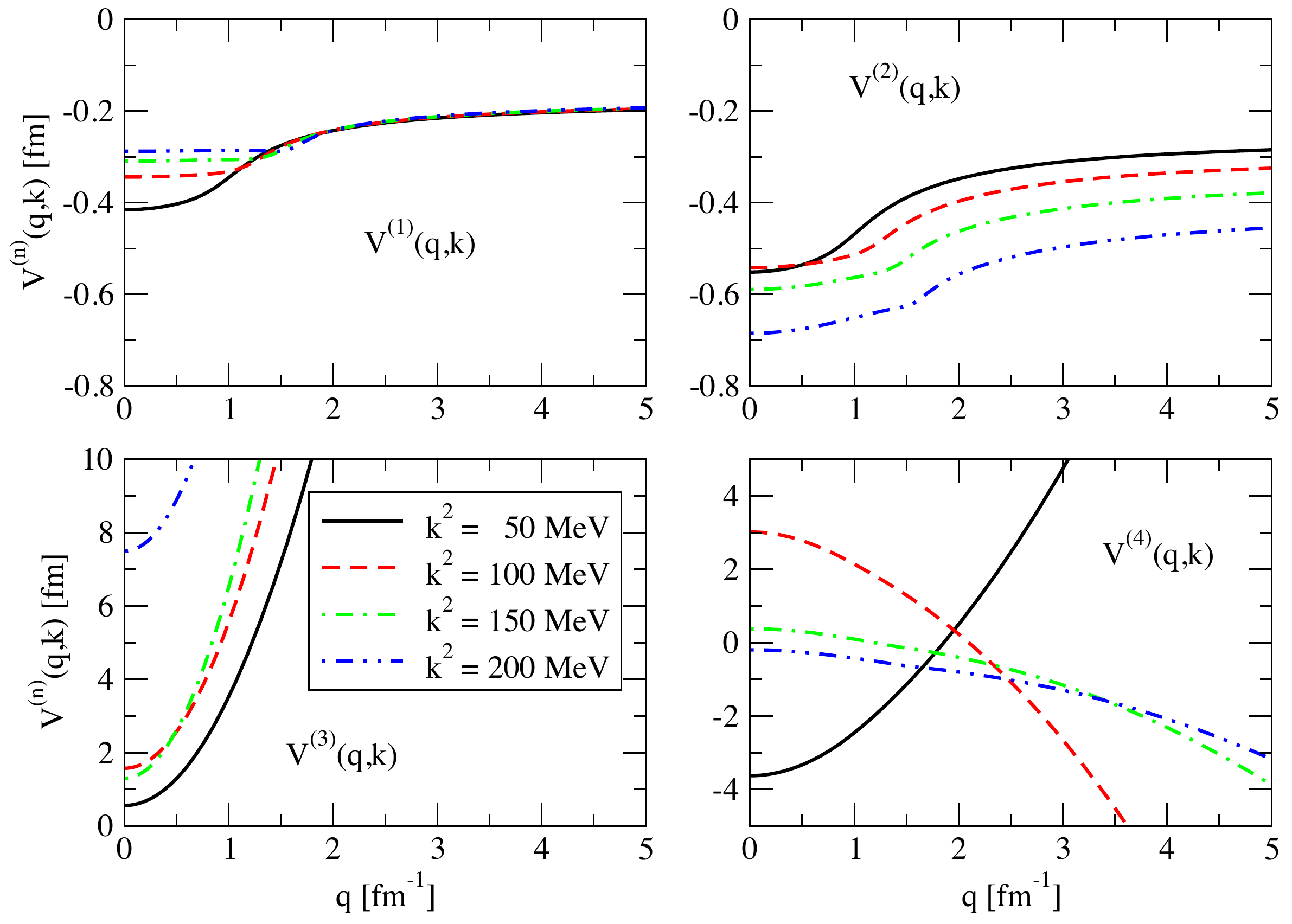}
\caption{(color online) Half-on-shell recursive driving terms
$V^{(n)}(q,k;k^2)$ for the $^1S_0$ channel,  corresponding to the
energies $E= (\hbar k)^2/m_N =$ 50, 100, 150 and 200 MeV
(with units such that $\hbar=1$ and $m_N=1$). The lines 
identification, shown inside the lower-left panel, are given for
all four panels. The values of the strengths, as well as the 
subtraction energy scales (not explicitly shown here), are  
given in Tables \ref{tab1} - \ref{tab4}.} \label{fig1}
\end{figure}
\begin{figure}[b]
\begin{center}
\includegraphics[height=5cm, width=8cm]{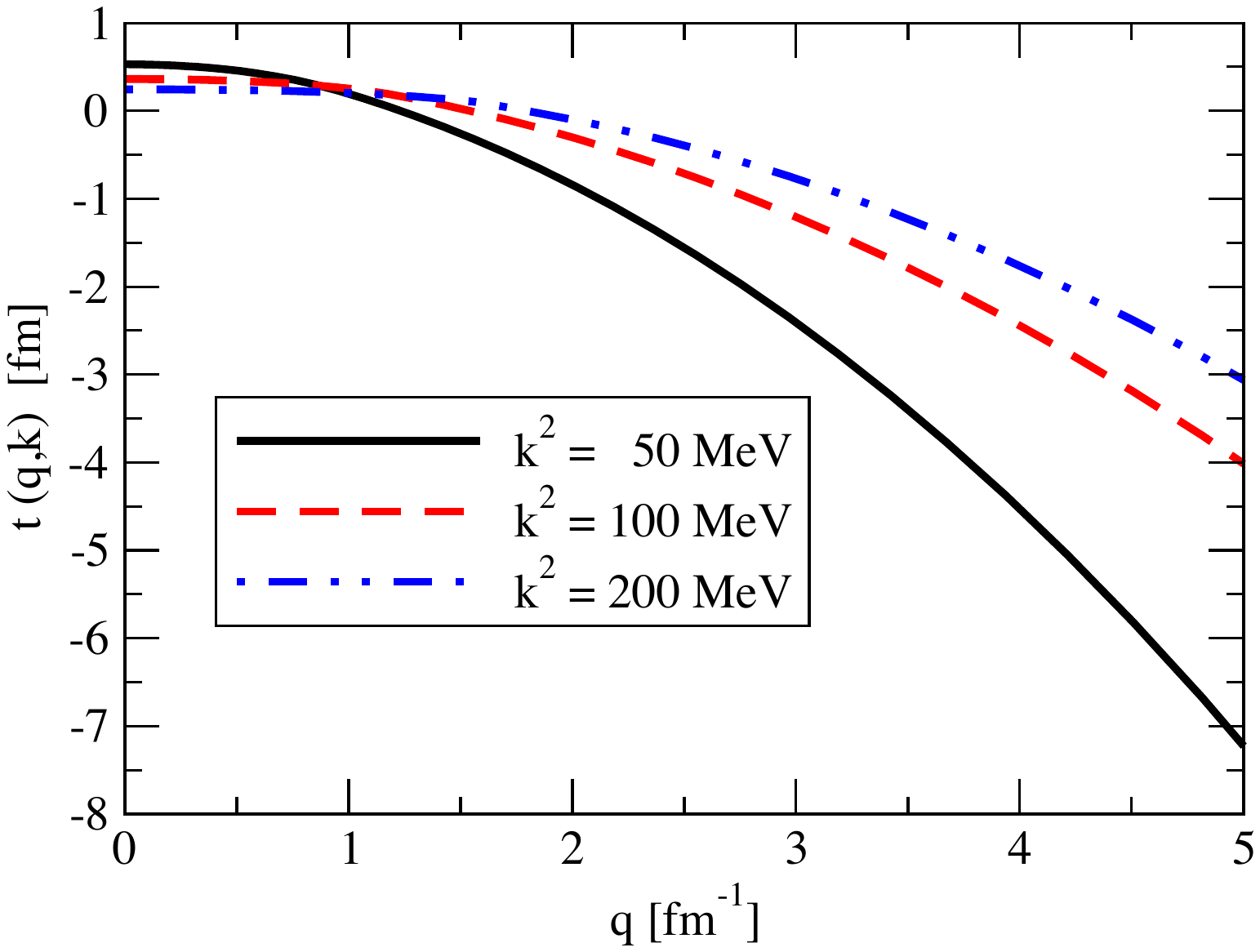}
\caption{(color online) Half-on-shell T-matrix elements for the $^1S_0$ channel, for the energies  $k^2=$ 50, 100 and 200 MeV.
As in Fig.~\ref{fig1}, the values of the strengths, as well as the 
subtraction energy scales (not explicitly shown here), are  
given in Tables \ref{tab1} - \ref{tab4}.}.
\label{fig2}
\end{center}
\end{figure}

\begin{figure}[t]
\centering
\includegraphics[height=10cm, width=12cm]{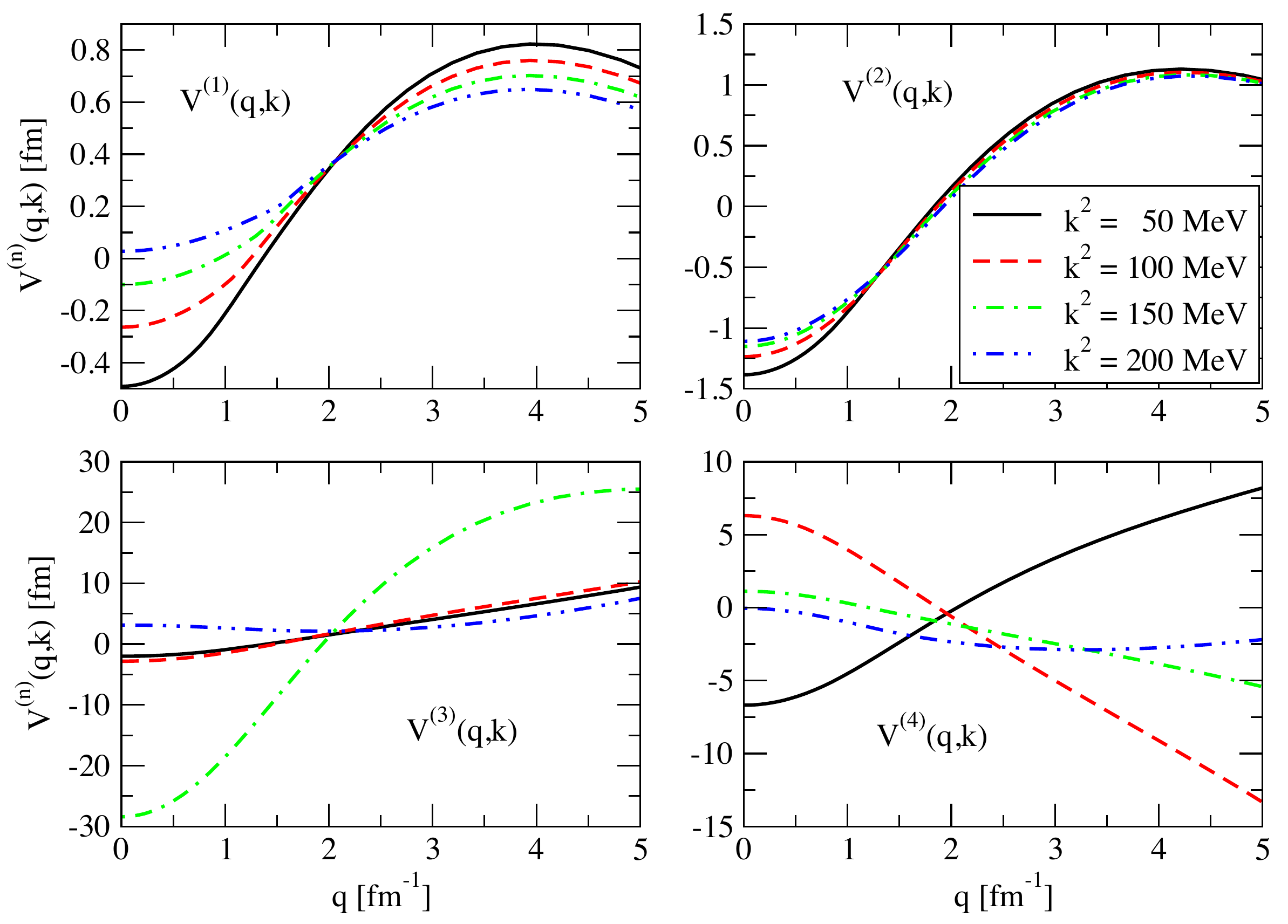}
\caption{(color online) Half-on-shell recursive driving terms $V^{(n)}(q,k;k^2)$
for the $^3S_1$ channel, corresponding to the energies $k^2 =$ 50, 100, 150
and 200 MeV, as given inside the upper-right panel.
As in Fig.~\ref{fig1}, the values of the strengths, as well as the 
subtraction energy scales (not explicitly shown here), are  
given in Tables \ref{tab1} - \ref{tab4}.}
\label{fig3}
\end{figure}
\begin{figure}[b]
\begin{center}
\includegraphics[height=5cm, width=8cm]{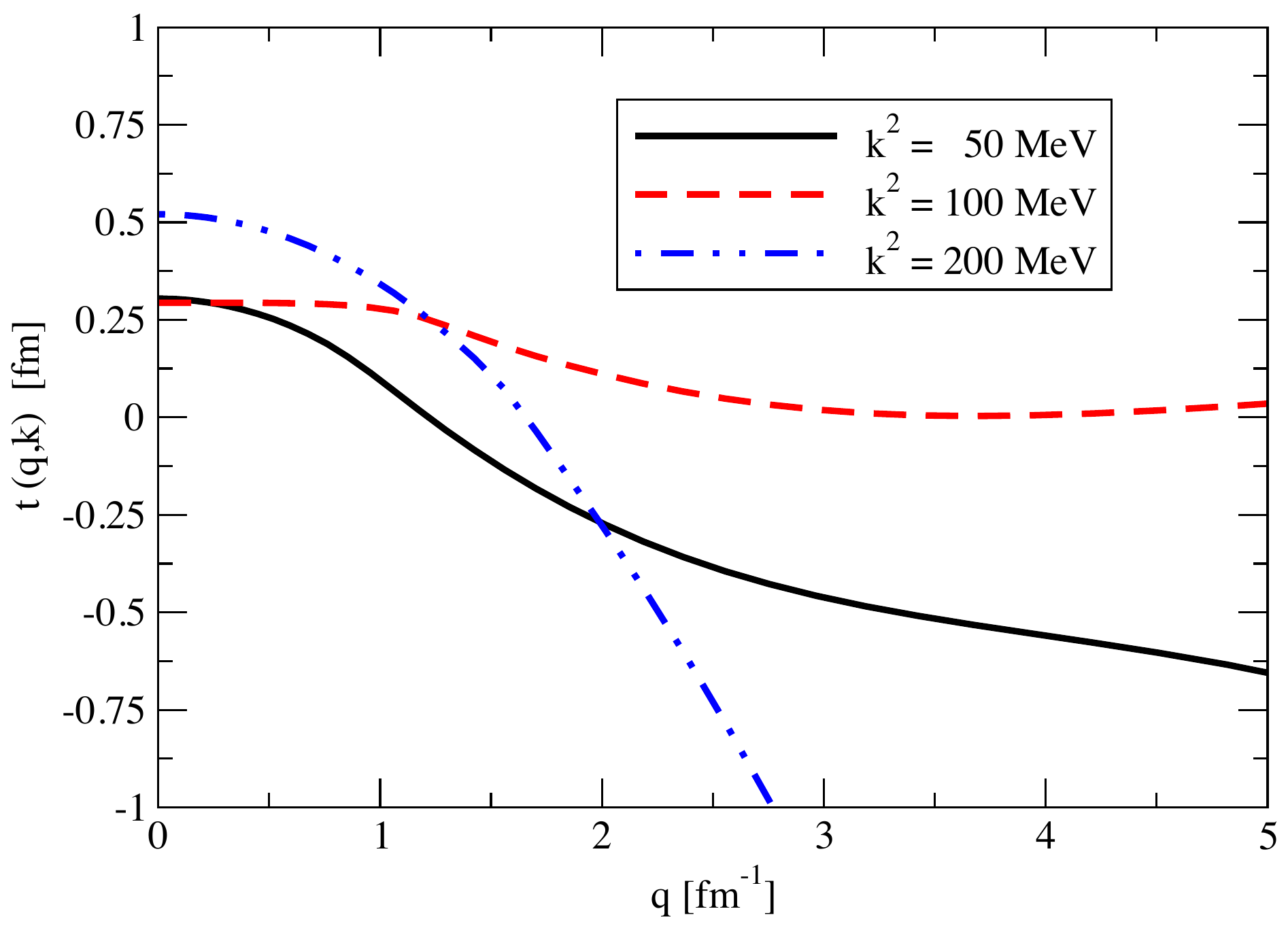}
\caption{(color online) Half-on-shell T-matrix elements for the $^3S_1$ channel,
for the energies $k^2 =$ 50, 100, 150 and 200 MeV. As in Fig.~\ref{fig1}, the values 
of the strengths, as well as the subtraction energy scales (not explicitly shown here), are  
given in Tables \ref{tab1} - \ref{tab4}.}
\label{fig4}
\end{center}
\end{figure}

We can observe the half-on-shell recursive driving terms evolving through the
four subtractions from $V^{(1)}(q,k)$ up to $t(q,k)$ for some values
of the on-shell momentum $k$ in Figs.~\ref{fig1} to  \ref{fig4}, for the
$^1S_0$ and $^3S_1$ channels. In Figs.~\ref{fig1} and  \ref{fig3} we have
the driving terms and half-shell T-matrix elements, respectively,
for the $^1S_0$ channel. Correspondingly, we have the results for the 
$^3S_1$ channel in Figs.~\ref{fig2} and  \ref{fig4}. These figures exhibit us an 
interesting finding due to Redish and Stickbauer~\cite{redish}.
They observe that distinct half-on-shell potentials may lead
to similar half-on-shell $T-$matrix. In other words, given
different $V(q,k)$ which fit the same on-shell observables their
corresponding half-on-shell $T-$matrix should be quite equivalent
despite the $V(q,k)$ discrepancies. That is why we observe a smooth
behavior of the scattering amplitude with  energy in 
Fig.~\ref{fig2} while the recursive driving terms, in
particular $V^{(4)}$, vary considerably as can be seen in the right
panel of Fig.~(\ref{fig1}).

The set of integral equations for the subtracted driving terms Eqs.
(\ref{v1ppl})-(\ref{v4+ppl}) given above also deserves further
comment to remove the naive misconception that the approach
given in \cite{plb05} ``involves invoking the Born approximation and consequently
unreliable for the higher singular potentials" \cite{YEP09}. As
one should observe, such approach relies on the definition of a
driving term $V^{(n)}$ at each subtraction order. Of course, such
term is the transition matrix at the subtraction point. The driving
term should be evolved dynamically by the renormalization procedure
to the next order, where higher divergent interactions are added, if
necessary. If a contact interaction is included, its strength is the
renormalized one at that point. Such procedure continues up to the
number of subtractions required to render finite the corresponding
scattering equation. Let us emphasize that, differently from the
usual Born approximation (where the dynamics does not evolve the
potential), this renormalization procedure relies on a recursive evolution
of the driving term, from one order to the next one. Moreover, it
produces a unitary $S-$matrix.

The LO term, OPEP plus Dirac-$\delta$ for $^3S_1-^3D_1$
and $^1S_0$ and OPEP for the higher partial waves are brought from
the fixed point at infinite to the reference scale $\mu^2$,
where the physical information is supplied to the two-nucleon
system. Moreover through the renormalization group equations we  are
able to evolve precisely the driving term of the subtracted
equations to any arbitrary scale without altering the physical
content of the observables. However, in this case the operator form
of $V^{(n)}$ will acquire a non-trivial form and will be not easily
identified with the starting potential.

\subsection{Full-on-shell amplitudes along the recursive renormalization process}

The next step is the computation of the full-on-shell matrices $V^{(n)}(k,k;k^2;-\mu^2)$,
obtained from the solution of the recursive Eqs. (\ref{v1ppl}) to (\ref{v4+ppl}), with $p=p'=k$.
Once we have the on-shell $T-$matrix, computed from Eq. (\ref{thon}) with $p=k$,
we can obtain the $S-$matrix as
\begin{eqnarray}
S &=& 1 - 2ik~T .\;
\end{eqnarray}

\begin{figure}[t]
\centering
\includegraphics[height=8cm, width=10cm]{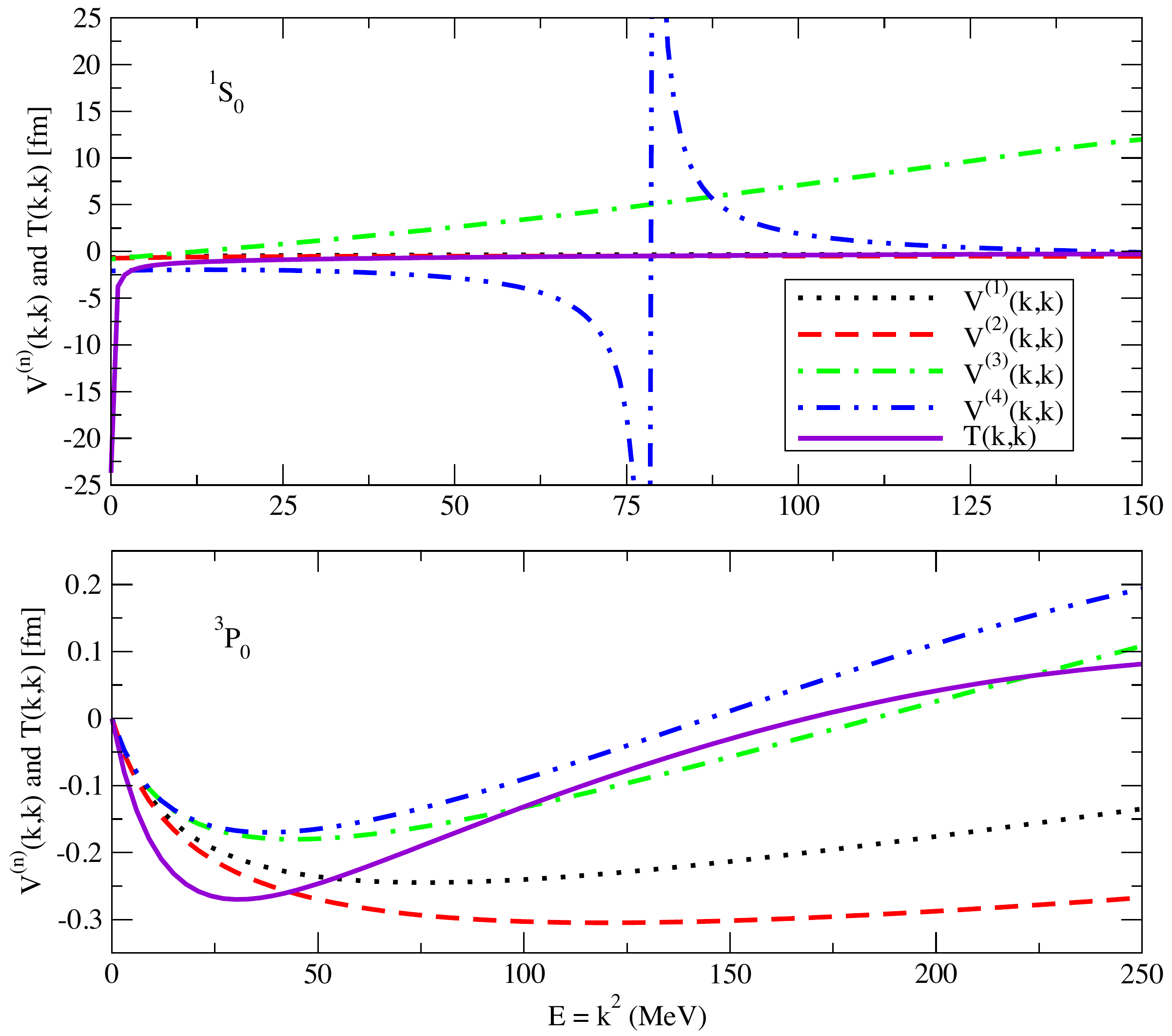}
\caption{(color online) On-shell elements $V^{(n)}(k,k;k^2)$ and $T(k,k;k^2)$ for the
uncoupled channels with $J=0$, as function of $k^2$.
The legends for the curves, given in the upper panel, are the same for both panels.
}
\label{fig5}
\end{figure}
\begin{figure}[b]
\centering
\includegraphics[height=8cm, width=10cm]{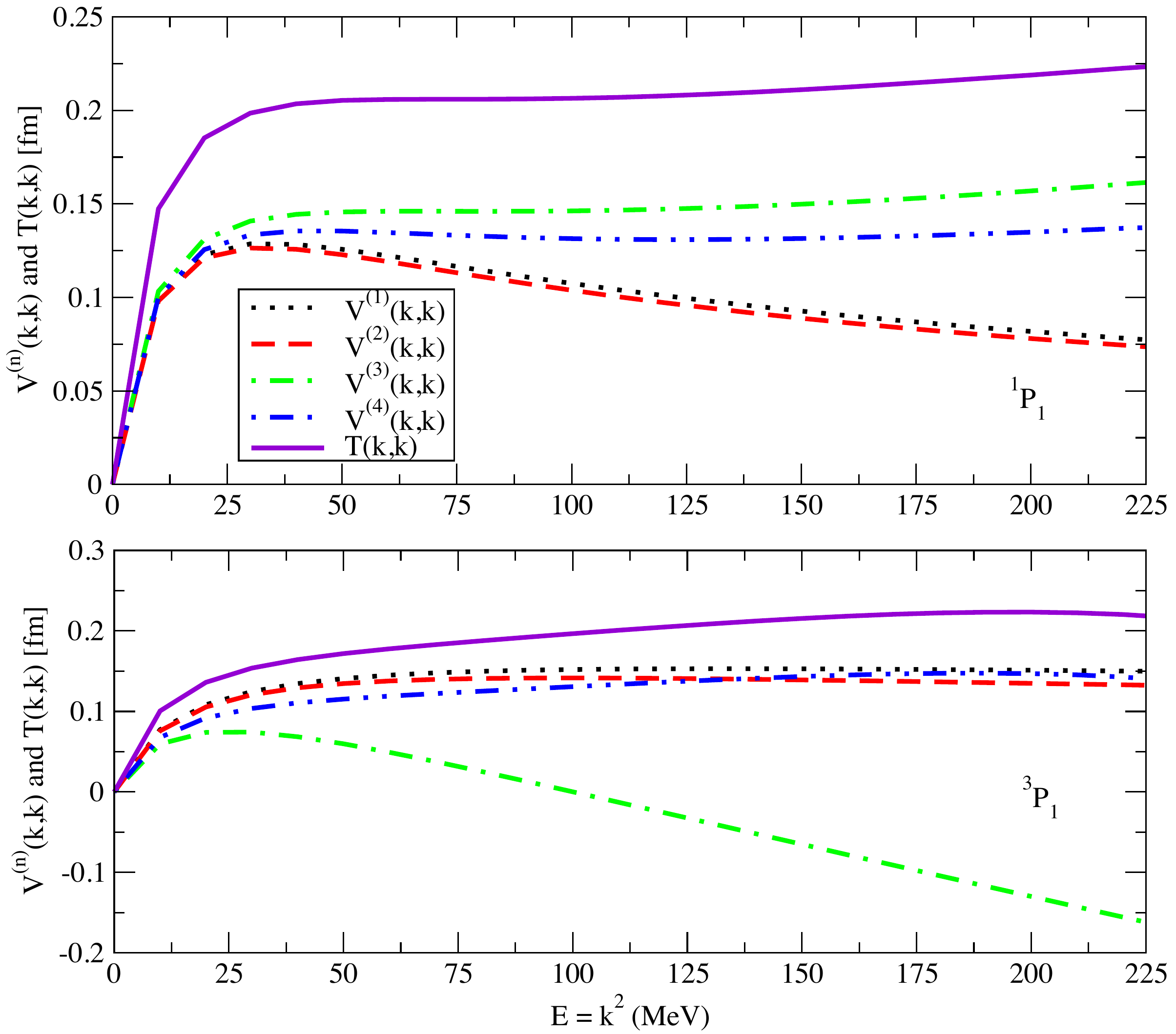}
\caption{(color online) On-shell elements $V^{(n)}(k,k;k^2)$ and $T(k,k;k^2)$ for the uncoupled channels with $J=1$, as function of $k^2$.
The legends for the curves, given in the upper panel, are the same for both panels.
}
\label{fig6}
\end{figure}

\begin{figure}[t]
\centering
\includegraphics[height=10cm, width=12cm]{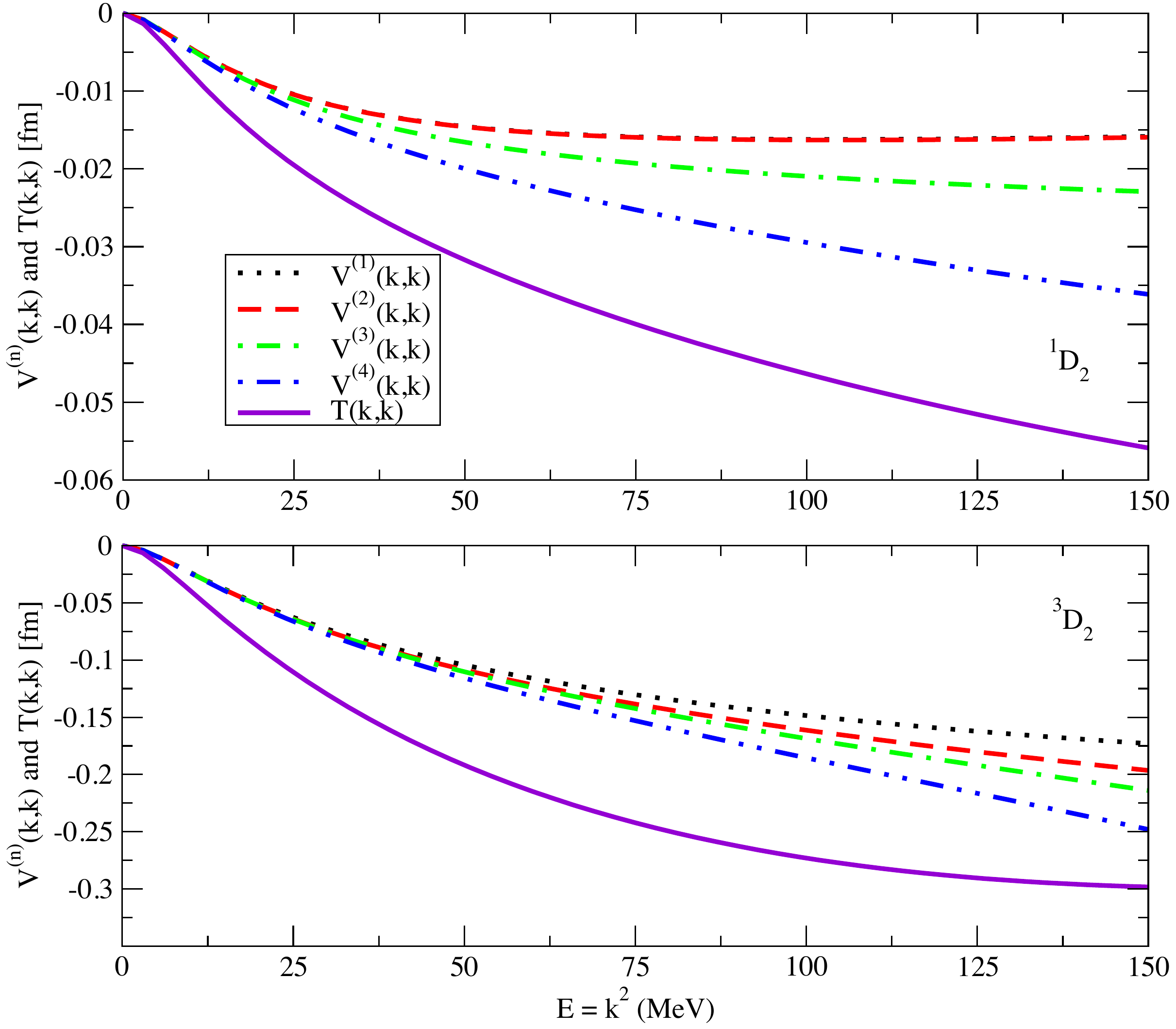}
\caption{(color online) On-shell elements $V^{(n)}(k,k;k^2)$ and $T(k,k;k^2)$ for the uncoupled channels with $J=2$, as function of $k^2$.
The legends for the curves, given in the upper panel, are the same for both panels.
}\label{fig7}
\end{figure}
\begin{figure}[t]
\centering
\includegraphics[height=12cm, width=12cm]{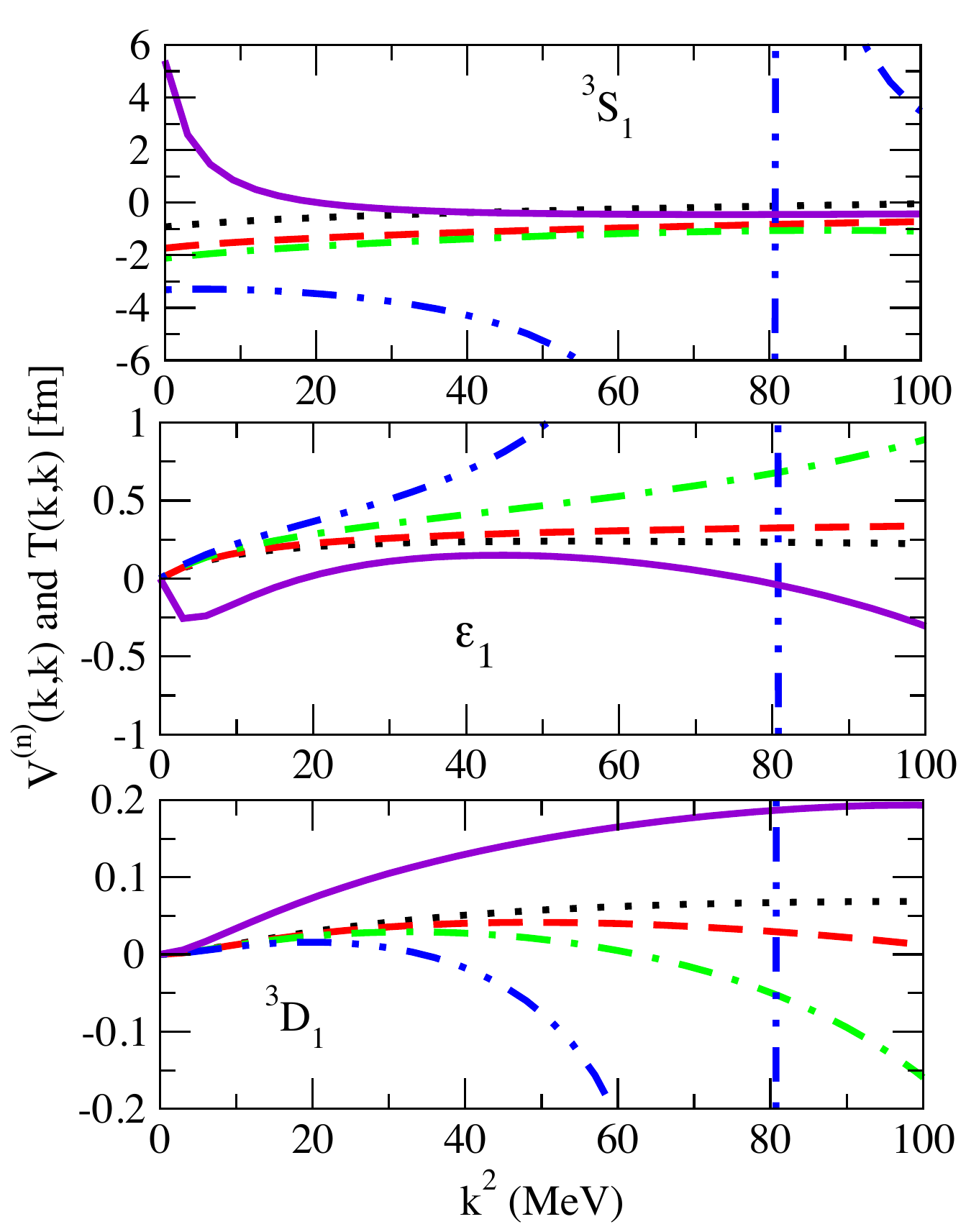}
\caption{(color online) On-shell elements $V^{(n)}(k,k;k^2)$ and $T(k,k;k^2)$ for the coupled channels with $J=1$. In the middle panel, the amplitude $T(k,k;k^2)$ for $\epsilon_1$ has been multiplied by a factor of $10$ in order to highlight its qualitative behavior. The legends for the curves appearing in the three panels are the same as the ones given in Fig.~\ref{fig7}.}
\label{fig8}
\end{figure}
\begin{figure}[t]
\centering
\includegraphics[height=12cm, width=12cm]{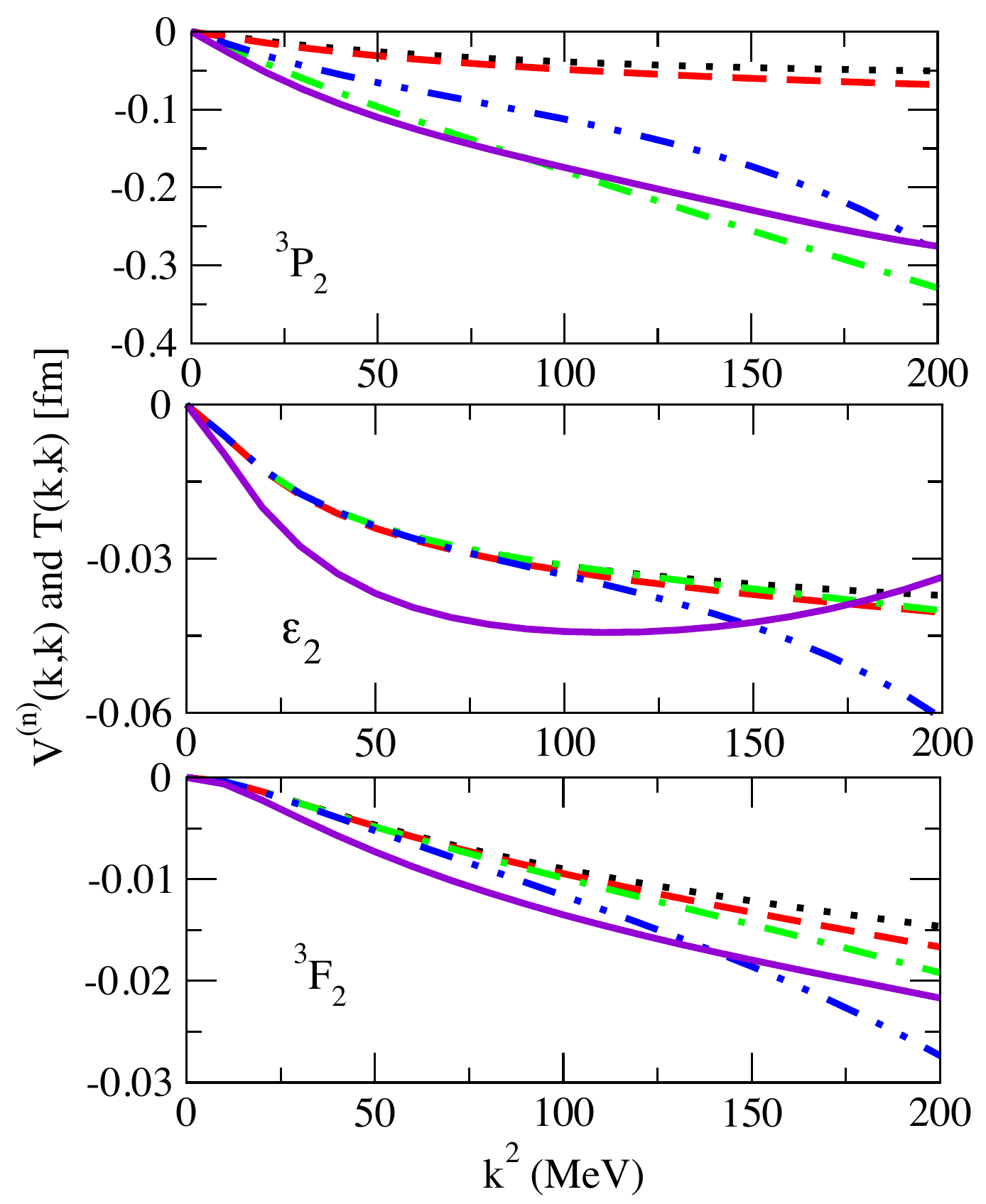}
\caption{(color online) On-shell  $V^{(n)}(k,k;k^2)$ and $T(k,k;k^2)$ for the coupled channels with $J=2$. The legends for the curves appearing in the three panels are the same as the ones given in Fig.~\ref{fig7}.}
\label{fig9}
\end{figure}

Figures \ref{fig5} - \ref{fig7} present the full-on-shell
recursive potentials, as we step forward in the recursive subtractive
renormalization process, and the real part of $T(k,k;k^2)$ (denoted
by $T(k,k)$ in the figures) in the uncoupled states with total
angular momentum up to $j=2$. For coupled channels, with $j=1$ and
$j=2$, the evolution of the recursive potential and the real part of
the scattering amplitude are shown in Figs.~\ref{fig8} and \ref{fig9}.

Results for $^1S_0$ and $^3P_0$ are shown in Fig.~\ref{fig5}.
Clearly the results of scattering amplitude for the singlet $S-$wave
channel are far from perturbative. At zero energy, it is observed
a strong deviation of $T(0,0;0)$ (that gives the scattering length) from
$V^{(4)}(0,0;0;-\mu^2)$, which is due to the nearby singlet virtual state.
The contribution of the contact in NLO to $V^{(3)}(k,k;k^2;-\mu^2)$
distinguishes it from the recursive process. In particular,
$V^{(4)}(k,k;k^2;-\mu^2)$ presents a pole, which arises from the
solution of the integral equation, that does not affect the scattering
amplitude in the energy range that is shown.
The comparison of $V^{(4)}(k,k;k^2;-\mu^2)$ with the real
part of $T(k,k;k^2)$ for $^3P_0$ shows a behavior that indicates the
dominance of the driven term.  The zero of both quantities are
close together and the magnitude of the real part of $T(k,k;k^2)$
decreases in respect to $V^{(4)}(k,k;k^2;-\mu^2)$ with energy. The
contribution of the NLO contact to $V^{(3)}(k,k;k^2;-\mu^2)$ is also
clearly seen in the figure, while the TPE NNLO potential appears to
be not so much relevant in this wave.

The on-shell matrix elements of the recursive potential for $^1P_1$
and $^3P_1$ and scattering amplitude are shown in Fig.~\ref{fig6}. 
For these waves the TPE NNLO potential does not gives
a relevant contribution as one observe from the  difference between
$V^{(4)}(k,k;k^2;-\mu^2)$ and $V^{(3)}(k,k;k^2;-\mu^2)$. The NLO
contact is important in these waves, as also seen for $^3P_0$. The
real part of scattering amplitude for $^1P_1$ is also somewhat
dominated by $V^{(4)}(k,k;k^2;-\mu^2)$. The uncoupled $D-$waves are
shown in Fig.~\ref{fig7}. No contacts are present and the
recursive potentials change smoothly from one to the next.

The coupled $^3S_1-^3D_1$ on-shell potentials and scattering
amplitudes are presented in Fig.~\ref{fig8}. The real part of
the scattering amplitudes for these channels are far from
perturbative. At zero energy  the real part of $t(0,0;0)$ for the
$^3S_1$ channel, that gives the scattering length, moves strongly
from $V^{(4)}(0,0;0;-\mu^2)$  due to the deuteron pole. The same is
observed at low energies for the off-diagonal amplitude, related to
the importance of the deuteron D/S ratio to the mixing parameter
\cite{mix}. The contribution of the contact in NLO to
$V^{(3)}(k,k;k^2;-\mu^2)$ distinguishes it from the recursive
process. The pole of $V^{(4)}(k,k;k^2;-\mu^2)$ does not affect the
scattering amplitude in the energy range of our calculations. The
coupled $^3P_2-^3F_2$ on-shell potentials and scattering amplitudes
are seen in Fig.~\ref{fig9}, and we observe a strong increase in
the magnitude of $V^{(4)}(k,k;k^2;-\mu^2)$, that also is an
indication of the contribution of the NNLO potential to this wave.

\section{Results for nucleon-nucleon phase-shifts and mixing parameters}

For the analysis of the phase-shifts and mixing parameters obtained
with the renormalized strengths and
subtraction energies presented in Tables \ref{tab1} to \ref{tab4}, we
adopt a systematics which splits the calculations in four sets:\\
{\it (i)} leading order (LO), as given in Eq.~(\ref{11});\\
{\it (ii)} leading order plus next-to-leading order contact
interactions
(LO + NLO CI), as given in Eq.(\ref{13});\\
{\it (iii)} full next-to-leading order (NLO), consisting of one-pion exchange,
two-pion exchange at NLO, and contact interactions;\\
{\it (iv)} next-to-next-to leading order (NNLO), which is the NLO plus
TPE diagrams at NNLO.\\
In particular, the set {\it (ii)} was inspired in the idea to promote some
NLO terms to LO, as a way to overcome difficulties with the Weinberg
power-counting rule (see Ref.~\cite{nogga05}). One
consequence of the failure of the naive dimensional analysis (NDA)
is that contact interactions that are subleading in Weinberg's power
conting are in fact necessary in order to make sense the T matrix
even at LO. In Refs.~\cite{wein-recipe,nogga05,va06}, it is shown how the origin
of the failure of NDA is the singularity of one pion exchange (OPE),
for which a single counterterm suffices at LO.

\begin{figure}[t]
\centering
\includegraphics[height=9cm, width=9cm]{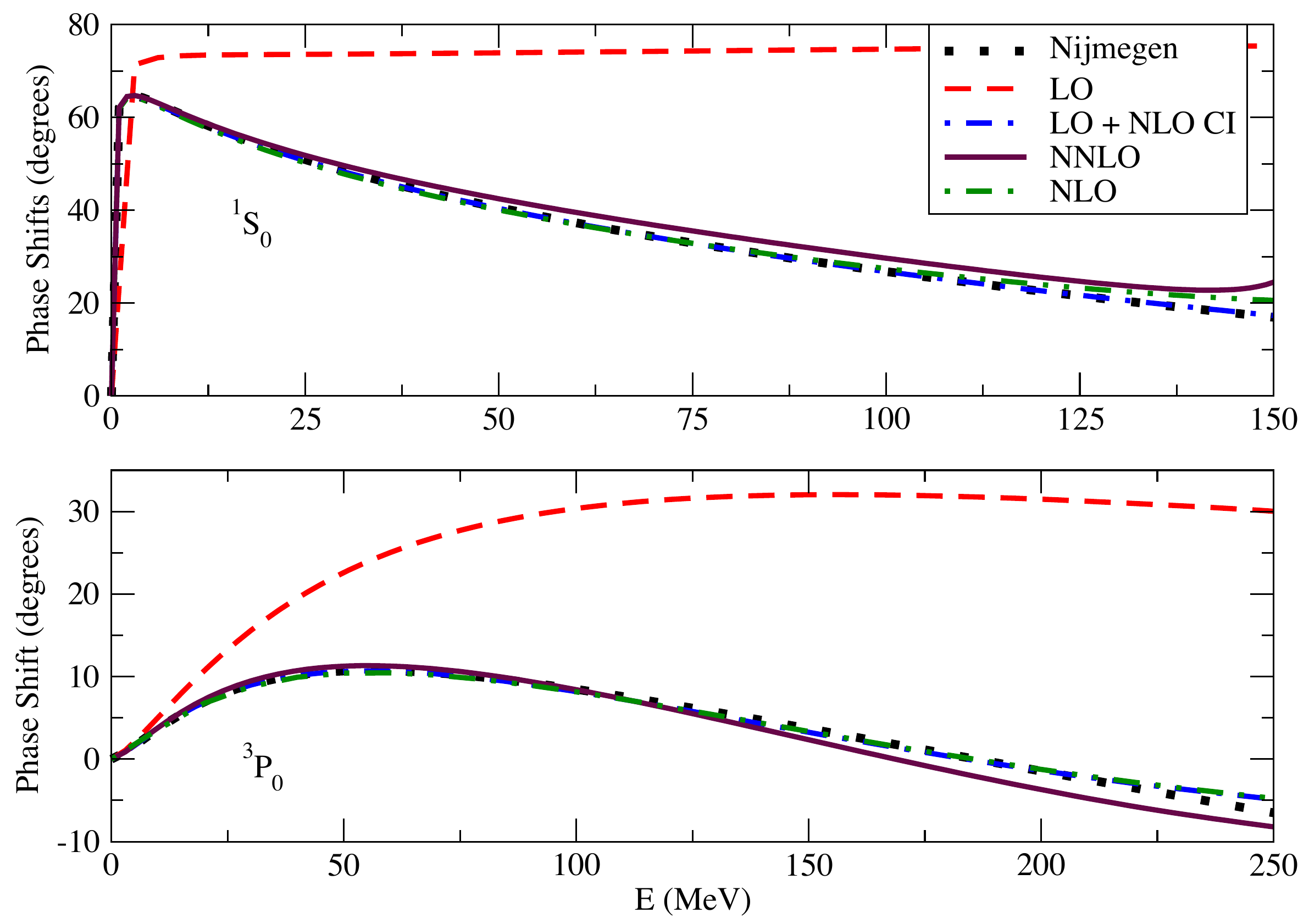}
\vspace{-.5cm}
\caption{(color online) Uncoupled channels for $J=0$.
Phase shifts for the $^1S_0$ wave with the subtraction point at -50 MeV, and for the $^3P_0$ wave
with the subtraction point at -100 MeV. The legends for the curves, given in the upper panel,
are the same for both panels.}
\label{fig10}
\vspace{-.5cm}
\end{figure}

For the LO potential in the singlet and triplet channels, the
respective scattering lengths are fitted. Indeed this calculation
reproduces the results obtained in Ref.~\cite{npa99}, which we
supply here for completeness and in order to compare with the
results obtained in NLO and NNLO. The results for the phase shifts
for the waves $^1S_0$ and $^3P_0$ are shown in Fig.~\ref{fig10}.
We should observe that results corresponding to the LO + NLO CI were
already presented in Ref.~\cite{plb05}. The NLO and NNLO
calculations were done by using the same renormalized strengths for
the derivative of the contacts (see $\lambda_2$ and $\lambda_3$ in
Tables \ref{tab2} - \ref{tab4}) and only refitted the singlet scattering
length by changing slightly $\lambda_0$. This means that at the
range of energies we perform our calculations the contributions of
TPE potentials in this wave are small. Moreover, the NLO and NNLO
calculations present a small systematic deviation steadily
increasing with energy. This is possibly due to the strong
attraction of the corresponding TPE NLO and NNLO potentials which
increases at higher momentum.

\begin{figure}[b]
\centering
\includegraphics[height=9cm, width=9cm]{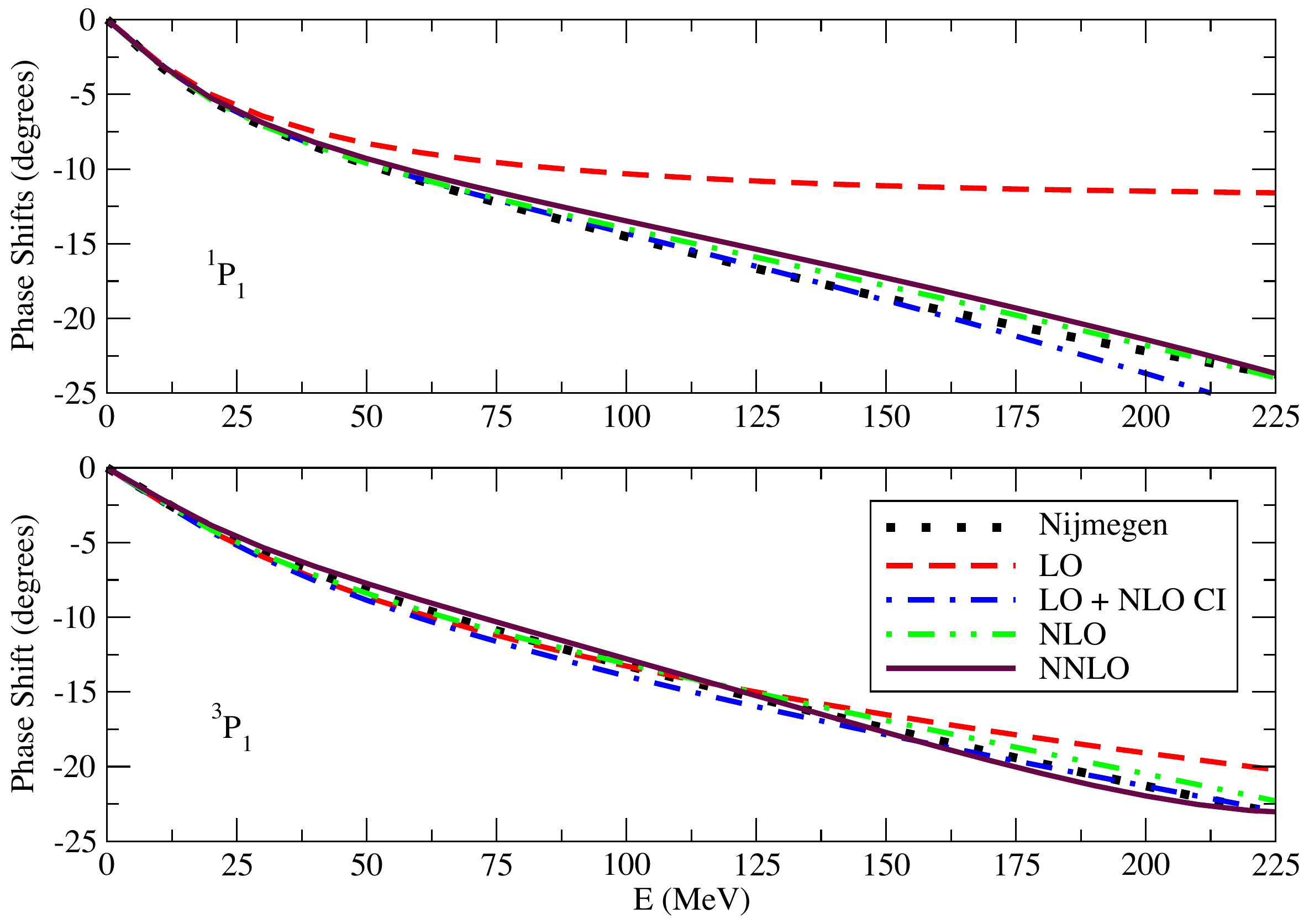}
\vspace{-.5cm}
\caption{(color online) Phase shifts for the $^1P_1$ and $^3P_1$ waves.
The legends for the curves are given in the lower frame for both the cases.}
\label{fig11}
\end{figure}

\begin{figure}[t]
\centering
\includegraphics[height=12cm, width=12cm]{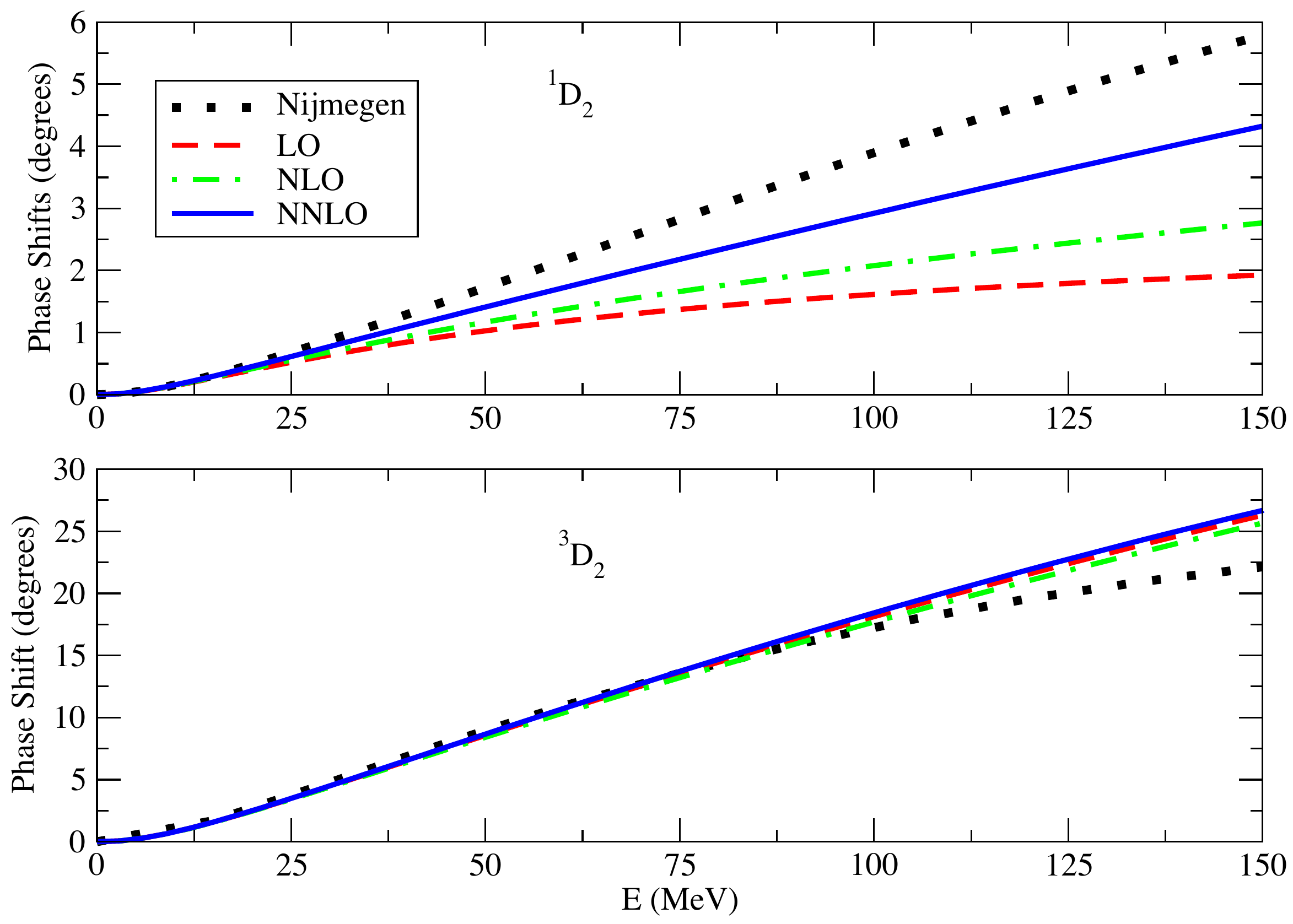}
\vskip 0.cm
\caption{(color online) Phase shifts for the $^1D_2$ and $^3D_2$ waves.
The legends for the curves are given in the upper frame for both the cases.}
\label{fig12}
\end{figure}

\begin{figure}[t]
\centering
\includegraphics[height=12cm, width=12cm]{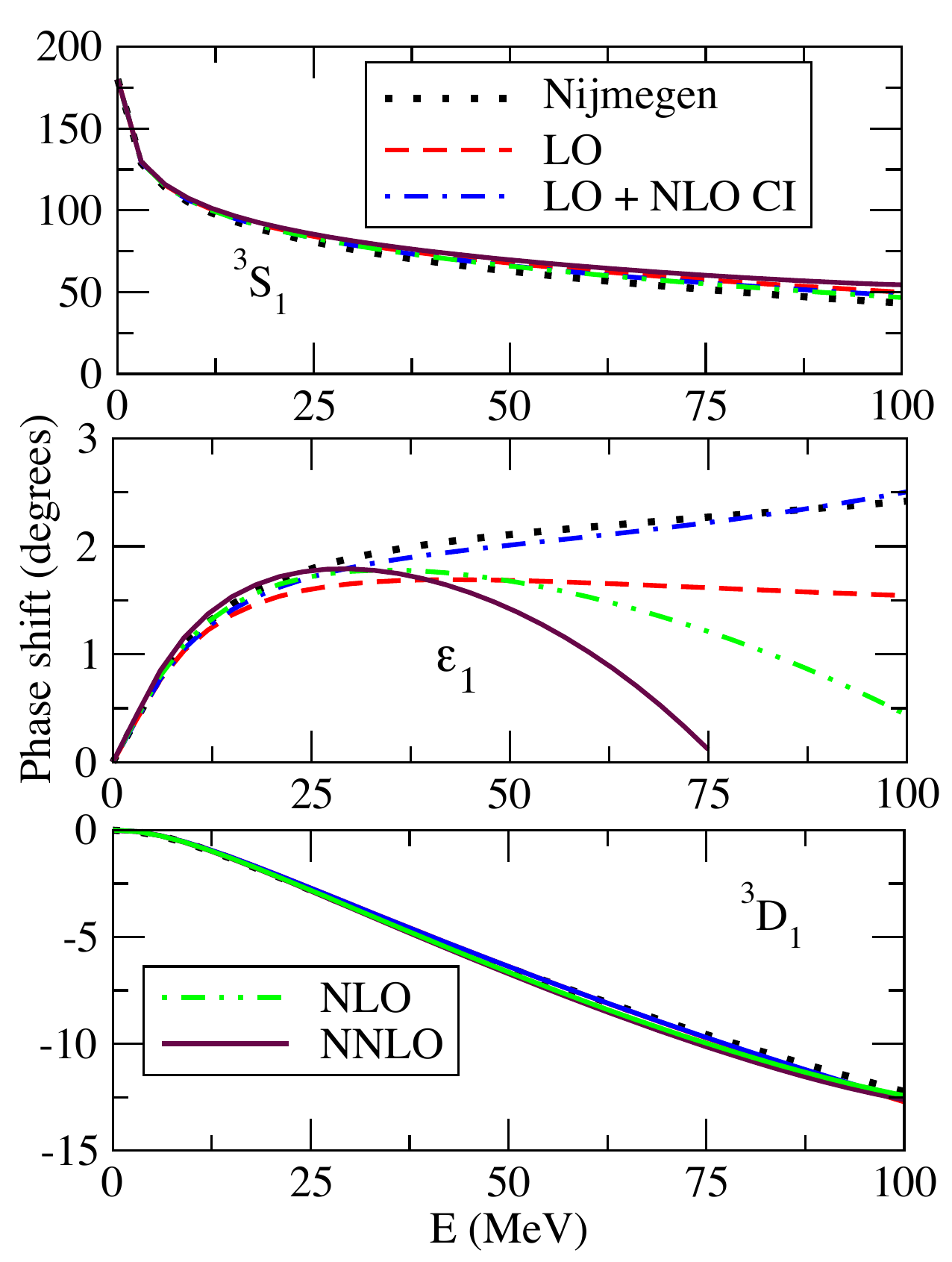}
\caption{(color online) Phase shifts and mixing parameter $\epsilon_1$
for the $^3S_1-^3D_1$ coupled channels. The legends for the curves appearing in the
three panels are shown in the upper and lower frames.}
\label{fig13}
\end{figure}
\begin{figure}[t]
\centering
\includegraphics[height=12cm, width=12cm]{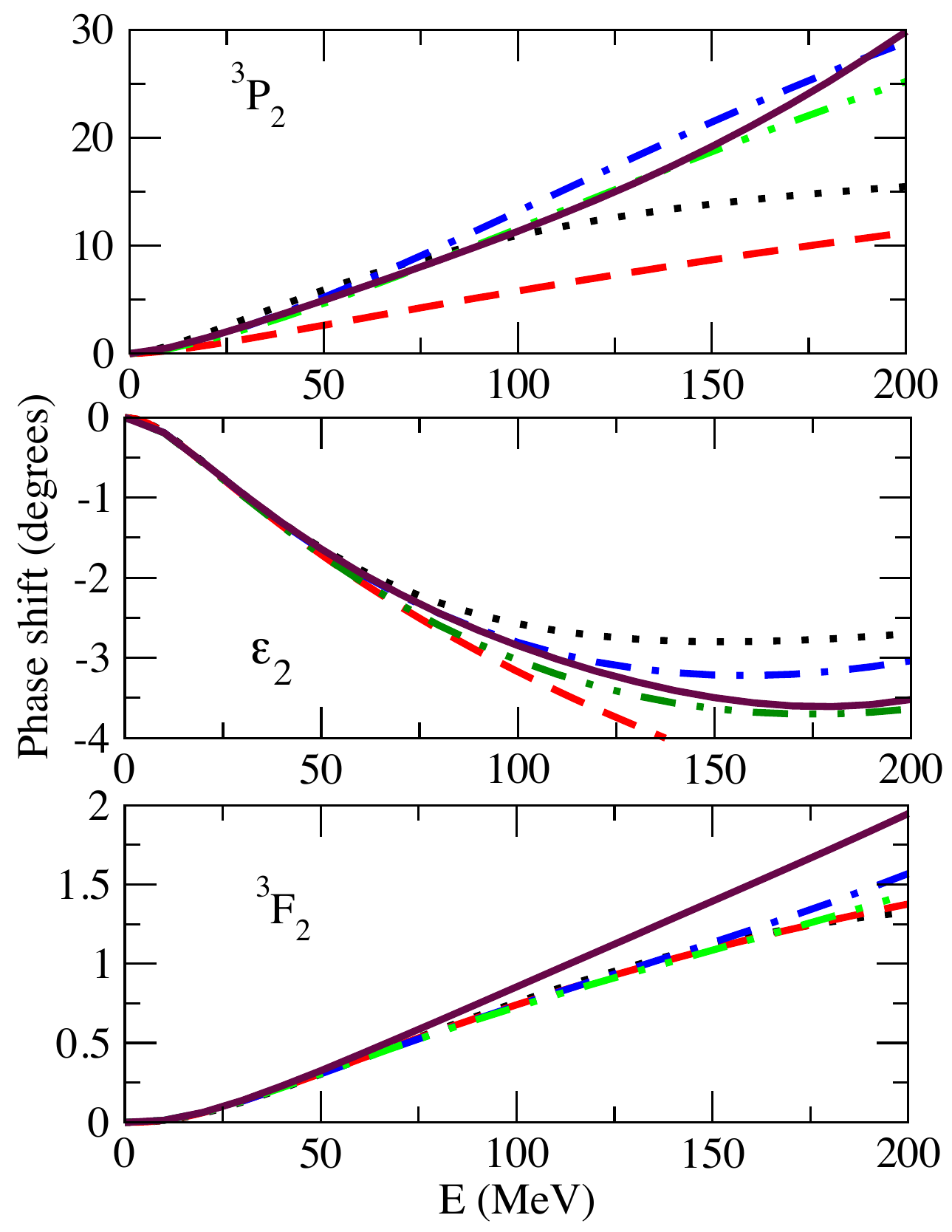}
\caption{(color online) Phase shifts and mixing parameter
$\epsilon_2$ for the $^3P_2-^3F_2$ coupled channels. The legends for the curves
appearing in the three panels are the same as the ones given in Fig.~\ref{fig13}.} \label{fig14}
\end{figure}

In the $^3P_0$ shown in the lower frame of Fig.~\ref{fig10} there
is a contact interaction of the type $V_{contact} = \lambda_1~p~p'$
at NLO along with one-pion and two-pion exchanges. The LO
calculation is not able to fit the scattering volume of $^3P_0$,
while with $\lambda_1$ given in Tables \ref{tab2} to \ref{tab4} we obtain a
better fit as shown by the corresponding adjustment of the
low-energy Nijmegen phase-shifts. It is seen that a small change in
$\lambda_1$ is enough to give the scattering volume for sets {\it
(ii)}, {\it (iii)} and {\it (iv)}. The small change in the contact
term which fits the singlet scattering length is similar to what is
found for $^3P_0$ in respect to the scattering volume.

The phase shifts for the waves $^1P_1$ and $^3P_1$ are shown in
Fig.~\ref{fig11}. In both channels there is a contact
interaction of the type $V_{contact} = \lambda_1~p~p'$ at NLO along
with one-pion and two-pion exchanges. In this case, we observe that
the LO + NLO CI is already giving a quite good fit, when comparing
with the Nijmegen results. The TPEP contribution is marginal as
observed by the slight change of $\lambda_1$ required to keep the
fit. (See Tables \ref{tab2} to \ref{tab4}).

The results for the phase shifts, for the uncoupled channels $^1D_2$ and $^3D_2$,
are shown in Fig.~\ref{fig12}.  As already mentioned before, no contact
terms are presented, in this case.

In the following, we discuss the results for the coupled channels for the
spin triplet $j=1$ in the $^3S_1-^3D_1$ states and $j=2$ in the
$^3P_2-^3F_2$ states. The phase shifts for the $^3S_1-^3D_1$ coupled
channels are shown in Fig.~\ref{fig13} and for  $^3P_2-^3F_2$ in 
Fig.~\ref{fig14}.

For the mixing parameter $^3S_1-^3D_1$, we consider $\epsilon_1$, as
defined in Ref.~\cite{stapp}, instead of the Blatt-Biedenharn
definition, $\epsilon_{BB}$~\cite{epsbb}. The reason for this choice
is related to the precise measurements available for $\epsilon_1$.
We observe in Fig.~\ref{fig13} that the mixing parameter $\epsilon_1$
can be well fitted with only the leading order plus a
small contact term in the mixed states (see $\lambda_4$ in Table \ref{tab2}). 
This indicates that the physics of $\epsilon_1$ seems well
controlled by OPEP, as long time ago predicted by a
shape-independent expansion~\cite{mix}.

What do we see when the NLO and NNLO potentials is inserted in our
method? The aforementioned nice fitting of $\epsilon_1$ by the LO
potential plus small contact term in the mixed channel disappears
[see Fig.~\ref{fig13}]. We met the widely recognized difficulty
that the effective potential has problems in the describing
$\epsilon_1$. To make concrete this point we mention that different
renormalization approaches, coordinate space renormalization method
\cite{variolarev} and subtraction plus cutoff \cite{YEP09} also
faced the same difficulty to fit $\epsilon_1$ in NLO and NNLO. In
particular Ref. \cite{YEP09} exploited the strong momentum cutoff
dependence to fit $\epsilon_1$, however the fit is not robust in the
sense that it should be smoothly cutoff dependent.
This common difficulties in different and independent calculations
reveals the distinct role played by the singularities in different
waves.

From the results shown in Fig.~\ref{fig13}, we see that the TPEP
does not exhibit a  systematic behavior in the different partial
waves regarding the NLO and NNLO potentials. From our point of view
a systematic behavior would require a cutoff for TPEP or its
inclusion only after higher order singular terms are included. This
problem is acute in the mixing parameter of the coupled channel
$^3S_1-^3D_1$. The mixing parameter with OPEP plus NLO contact is
well fitted up to 100 MeV lab energy. The introduction of NLO and
NNLO TPEP, clearly worsen the fit to the Nijmegen phase-shift
analysis~\cite{nij}. To fit the mixing parameter it is important to
have the deuteron asymptotic D/S ratio within their accepted value.
This was pointed out in a recent work~\cite{variola09}, where the
authors have also included the contribution of the Delta excitation. 
Indeed, long-time ago, it was already concluded that the mixing 
parameter at low energies is determined by the deuteron properties 
and by OPEP~\cite{mix}. 
By considering the Blatt-Biedenharn~\cite{epsbb} definition, it 
was demonstrated in \cite{mix} that the correct long-range behavior of the 
tensor potential is essential for a realistic reproduction of the mixing 
parameter. As shown in \cite{mix}, separable tensor Yamaguchi and square-well 
potentials, which do not possess 
the OPE tail, when fitted to reproduce the deuteron binding and asymptotic 
normalization, badly fail to reproduce the correct value of the mixing 
parameter.
The NLO and NNLO potentials for
the deuteron channel seem to give a too strong contribution,
enhanced by the singular behavior of OPEP at short distances in this
channel. Although, the diagonal and off-diagonal NLO contact plus
OPEP and LO contact are enough to give a nice fitting to the mixing
parameter up to $E_{\text{lab.}} = 100~\text{MeV}$, the inclusion of
NLO and NNLO TPEP does not provide a good fit, and the results
systematically deviate from the Nijmegen data for $\epsilon_1$. This
indicates that by keeping the OPEP potential intact, the NLO and
NNLO TPEP potentials have to have a cut at short distances.

It is appropriate to summarize the comparison between  our
method of multiple subtractions and that used in ref. \cite{YEP09}.
In ours no cutoff is needed while in the later one just one
subtraction is required since higher order singularities coming from
the delta derivatives are tamed by a cutoff. Numerically,  in
general, both procedures lead to a similar fitting for the S and P
waves up to energies of about 200 MeV. We also remark that the
mixing parameter, $\epsilon_1$, presents the same deficiency in the
fit by both methods as well as for a different regularization and
renormalization method in coordinate space without the contribution
of the Delta excitation \cite{variola09}. 
\begin{figure}[t]
\centering
\includegraphics[height=8cm, width=12cm]{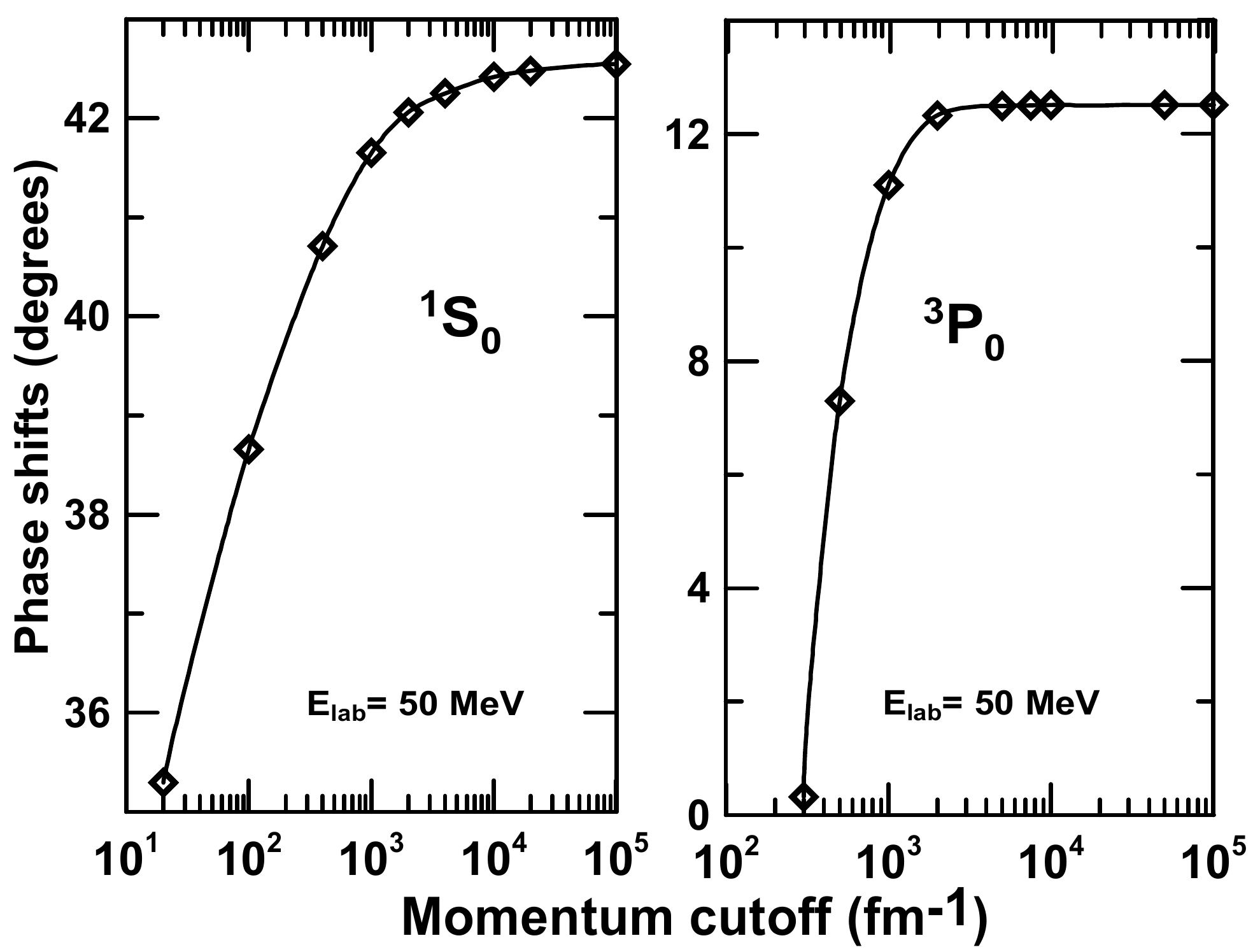}
\caption{
 Phase shifts for the $^1S_0$ (left panel) and $^3P_0$ 
(right panel) waves, for $E_{lab}=$ 50 MeV, exemplifying 
a general behavior of the phase shifts for increasing cutoffs.  
Both the cases presented here were calculated considering the 
NNLO potential, using the $n=$4 subtracted scattering equation 
(\ref{thon}), with parameters given in Table \ref{tab4}.
The same kind of behavior is verified for the other
orders of the potential (LO and NLO), as well as for other
waves and $E_{lab}$ energies.
}
\label{converg}
\end{figure}

We note that, once the parameters are fitted
(renormalized strengths and subtraction points) the subtraction
scale can be moved without changing the T-matrix results. Our method 
is RG invariant by construction. The flow equation (\ref{ss3}), which
transforms the driven term of the subtracted scattering equation
(\ref{thon}) under the dislocation of the subtraction point, was
derived from the invariance of the scattering matrix under
variations of $\mu$.

To conclude this section, we illustrate the cutoff independence of 
our method. We choose to present results for two singular cases,
the $^1S_0$ and $^3P_0$ phase-shifts at 50 MeV.
They were obtained by considering the NNLO potential, with $n=4$
subtracted scattering equation (\ref{thon}).
As shown in Fig.~\ref{converg}, the phase-shifts stabilize when the 
cutoffs are driven towards infinity. 
The same kind of cutoff behavior is verified for the other
orders of the potential (LO and NLO), as well as for other partial 
waves and $E_{lab}$ energies. Our results shown along this paper  
were obtained for infinite cutoffs, with the verified convergence 
due to the subtractive renormalization approach we are considering.

\section{Conclusion}

We present a systematic methodology to renormalize the
nucleon-nucleon interaction using a recursive subtracted approach
with multiple subtractions in the kernel. Within the subtractive
scheme we studied the two-nucleon scattering $T-$matrix for the NLO
and NNLO potentials, for all partial waves up to $j=2$. The
renormalized strengths of the contact interactions, the so-called low-energy
constants, were fitted to the low-energy phase-shifts and mixing parameters
from the Nijmegen partial wave analysis for a reference subtraction point.

In order to show how the multiple subtraction renormalization scheme
can be implemented, an analytical example is given for the $P-$wave
channel. In that case, two subtraction are required to eliminate the
ultraviolet singularity of the interaction. We give explicitly the
solution of the RG equation. Although we have not fully explored
the analytical form of this $P-$wave amplitude, we call the
attention of the reader to the richness of the analytical
continuation to the complex energy plane in order to obtain virtual
and resonant states~\cite{roriz,bertu-bira,del-glo}.

{In a sharp or smooth cutoff approach, the LO, NLO and NNLO
potentials  are regularized to vanish above a certain momentum scale
and then inserted in the LS equation. In our method, the cutoff is
purely instrumental and the limit of the momentum cutoff going to
infinity can be easily performed, since a finite $T-$matrix is ensured
by  multiple subtractions in the scattering equation. The original
potential is kept intact and enters in the recursive process as we
described in detail.}

We analyze the matrix elements of the potentials in momentum space
for the $^1S_0$ channel as we go from LO through NLO up to NNLO.
This suggest in a practical way a momentum scale, which we associate
with the subtraction point, where the systematic expansion of the
potential from the chiral effective field theory should be used
within our subtraction scheme. From such analysis, we verify that a
momentum scale of about 1 fm$^{-1}$ separates the matrix elements of
NLO and NNLO potentials in two regions: below this scale they are
comparable to the OPEP. With this indication, we choose a
subtraction point at an energy $-\mu^2\sim$ -50 to -100 MeV.
 The subtraction momentum scale comes to be at the order of the
QCD scale, $\Lambda_{QCD}$\cite{pdg} and well below the $\rho-$meson
mass, consistent with the general suggestion of Weinberg\cite{wei}.
Important to observe that, by taking into account the values of the NN
low energy parameters and the form of the bare interaction including 
the derivatives of the contacts, a renormalization scale near  
the value of the $\Lambda_{QCD}$ is consistent with the Wigner 
bound~\cite{1997phillips,2011szpigel}. Once we fit the renormalized strengths 
of the contacts, considering the pion exchange potentials at LO, NLO and
NNLO, the subtraction point can be moved arbitrarily. Our method is
RG invariant by construction, and the flow equation, (\ref{ss3}),
transforms the driven term of the subtracted scattering equation
such that scattering matrix invariant under dislocation of $\mu$.

We show how the half-on-shell potentials for $^1S_0$ and $^3S_1$ channels
evolve through the four subtractions, from $V^{(1)}(q,k)$ up to $T(q,k;k^2)$.
These exhibit a relation with the interesting finding of
Redish-Stickbauer~\cite{redish} that the half-on-shell potential and
corresponding half-on-shell $T-$matrix can be very different. Given
distinct $V(q,k)$, which fit the same on-shell observables, the
corresponding half-on-shell $T-$matrix should be quite equivalent in
spite the sharp $V(q,k)$ differences. That is consistent with the
smooth behavior of the scattering amplitude with energy, while the
recursive driving terms can vary considerably.

In the $^3P_0$ channel, a contact interaction is introduced at NLO
together with OPEP and TPEP. The LO calculation is not able to fit
the scattering volume of $^3P_0$, while the addition of contact
gives a better fit of the Nijmegen phase-shifts. It is verified that
a small change in the renormalized strength of the contact, obtained
only with OPEP, is enough to reproduce the scattering volume for
NLO and NNLO. The same behavior is observed for the strength of the
contact interaction when the singlet scattering length is fitted
with NLO and NNLO potentials.

For the $^1P_1$ and $^3P_1$ channels we introduce a contact
interaction at NLO along with OPEP and TPEP. A quite good fit of the
Nijmegen phase shifts is obtained with OPEP plus the contact.
Again we observe that the TPEP contribution is marginal as a
 slight change of the renormalized strengths keeps the
fit. These observations indicate the absence of an essential
singularity in the OPEP and TPEP potentials up to NNLO in uncoupled
$P-$waves, beyond the contacts itself, differently from what is
found for the coupled $^3S_1-^3D_1$ channel.

After our fitting of the contact interactions for the $P-$waves
($^3P_0$, $^3P_2$, $^1P_1$, $^3P_1$), the following observations
apply to the matrix elements in momentum space of the LO, NLO and
NNLO potentials. For $^3P_0$ and $^1P_1$, the contact dominates
above 3 fm$^{-1}$, while below 2 fm$^{-1}$ the LO potential
dominates. For the $^3P_2$, the LO potential is weak, while the
inclusion of the contact interaction enhances the attraction of the
NLO potential, which weakens the strong  repulsion from the NNLO
TPEP.

In the next, we summarize our findings for the coupled channels:
$^3S_1-^3D_1$ and $^3P_2-^3F_2$.
The mixing parameter $\epsilon_1$ can be well fitted at LO
plus a small contact term which couples the $S-D$ states in the
interaction. This indicates that the physics of $\epsilon_1$ seems
well controlled by OPEP, as suggested in \cite{mix} by a
shape-independent expansion.
When we include the NLO and NNLO potentials within our
method, the nice fitting $\epsilon_1$ is destroyed.
The NLO and NNLO potentials for the deuteron channel seems to give
a too strong contribution, enhanced by the singular behaviour of
the OPEP at short distances in this channel.

The difficulty in fitting  $\epsilon_1$ with effective
potentials in NLO and NNLO, is recognized by different
renormalization approaches~\cite{variolarev,YEP09}.
Such shortcoming is related to the strong momentum cutoff
dependence of $\epsilon_1$, as found for example by the
coordinate renormalization approach~\cite{variolarev}, possibly
due to the singular behavior of OPEP~\cite{bira2008}.

The contribution of TPEP does not exhibit a systematic behavior in
the different partial waves regarding the NLO and NNLO potentials.
Our results suggest that, for a systematic behavior, the TPEP should
be weakened for momentum larger than few fm$^{-1}$'s. This could be
done via a cutoff or by considering higher order contacts, which
could suppress the TPEP contribution in the appropriate partial
waves.

We found that the derivative contact terms dominate over the NLO and
NNLO two-pion exchange interactions in the $S-$wave channels,
starting at low-momentum scales ($\sim$ 0.3 fm$^{-1}$). In $P-$wave
channels the contacts are also important for the fitting of the
corresponding phase-shifts.

Finally, we should observe that the input of our method is the
$T-$matrix at a given energy, where the physical information is
supplied to the two-nucleon system. When the interaction is at LO,
the point where the physical input is given is not constrained. Once
we move to NLO and NNLO, the energy $-\mu^2$ arises as a scale where
the low-energy observables can be obtained. In view of that, the
particular value of the subtraction point acquires the status of a
physical scale when the NLO and NNLO interactions are introduced. It
is gratifying to verify that our fittings with the associated
subtraction point, given by the renormalization scale, comes to be
about $\Lambda_{QCD}$, well below the $\rho-$meson mass. This is
consistent with the general Weinberg's\cite{wei} suggestion, that an
effective potential should be valid for momenta much smaller than a
typical QCD scale of 1 GeV, and the intermediate nucleon-nucleon
propagation should be damped at such small momentum scale.

\section*{ACKNOWLEDGMENTS}
We thank the Brazilian agencies Funda\c c\~ao de Amparo \`a Pesquisa
do Estado de S\~ao Paulo (FAPESP) and Conselho Nacional de Desenvolvimento
Cient\'\i fico e Tenol\'ogico (CNPq) for partial support.
V.S.T. would like to thank FAEPEX/UNICAMP for partial support.

\appendix

\section{NN Phase Shifts and mixing parameters}

We follow the definitions given in Ref.~\cite{stapp} for the nucleon-nucleon phase shifts and mixing parameters, which are appropriate to the case that we have coupled channels.

Considering the cases with $j>0$, for the coupled channels we have an 
$S-$matrix expression given by
\begin{equation}
(S)_{j} \equiv \left(
\begin{array}{cc}
S_{j-1,j-1} & S_{j-1,j+1}  \\
S_{j+1,j-1} & S_{j+1,j+1}
\end{array}
\right) 
=\left(
\begin{array}{cc}
e^{{\rm i}\delta_{j-1}} & 0  \\
0 & e^{{\rm i}\delta_{j+1}}
\end{array}
\right)
\left(
\begin{array}{cc}
\cos(2\epsilon_{j}) & {\rm i}\sin(2\epsilon_{j})  \\
{\rm i}\sin(2\epsilon_{j}) & \cos(2\epsilon_{j})
\end{array}
\right)
\left(
\begin{array}{cc}
e^{{\rm i}\delta_{j-1}} & 0  \\
0 & e^{{\rm i}\delta_{j+1}}
\end{array}
\right),
\end{equation}
from where the corresponding phase shifts and mixing parameter are 
\begin{equation}
\delta_{j-1} = \frac 1 2 \tan^{-1}
\left(\frac{Im~S_{j-1,j-1}}{Re~S_{j-1,j-1}}\right) \; ,
\end{equation}

\begin{equation}
\delta_{j+1} = \frac 1 2 \tan^{-1}
\left(\frac{Im~S_{j+1,j+1}}{Re~S_{j+1,j+1}}\right) \; ,
\end{equation}

\begin{equation}
\epsilon_j = -\frac 1 2 \tan^{-1}\left( \frac{{\rm i}~(S_{j-1,j+1}+
S_{j+1,j-1})}{2\sqrt{S_{j-1,j-1} S_{j+1,j+1}}}\right) \; .
\end{equation}

\end{document}